\documentclass[aps,prl,superscriptaddress,twocolumn,showpacs,longbibliography,nofootinbib]{revtex4-1}
\PassOptionsToPackage{usenames,dvipsnames}{color}
\usepackage{amsmath,amsfonts,amssymb,amsthm,graphicx,bbm,enumerate,times,color,mathrsfs}
\usepackage{dsfont}
\usepackage{times}
\usepackage[normalem]{ulem}
\usepackage{mathtools}
\usepackage[makeroom]{cancel}

\usepackage{natbib}
\setcitestyle{square,comma,numbers}
 \usepackage{amsfonts}
\usepackage{amsmath}
 \usepackage{color}
 \usepackage{amssymb}			

\usepackage[colorlinks=true,citecolor=blue,linkcolor=blue,urlcolor=blue]{hyperref}

\usepackage{amsmath,amsfonts,amssymb}
\usepackage{mathtools}
\usepackage{simpler-wick}

\newcommand{\ket}[1]{|#1\rangle}
\newcommand{\bra}[1]{\langle #1|}
\newcommand{\braket}[2]{\langle #1|#2\rangle}
\newcommand{\Tr}{\mathord{\rm Tr}}
 
\newcommand{\eexp}[1]{\mathrm{e}^{#1}}

\def\oned{\mathrm{1d}}
\def\dd{\mathord{\rm d}}

\def\ii{\mathord{\rm i}}
\newcommand{\klamm}[1]{{(#1)}}
\newcommand{\multInt}{\int \hspace{-1mm}...\hspace{-1mm} \int}
\newcommand{\multIntbi}{\int_{-\infty}^\infty \hspace{-1mm}...\hspace{-1mm} \int_{-\infty}^\infty}
\newcommand{\multIntone}{\int_{0}^\infty \hspace{-1mm}...\hspace{-1mm} \int_{0}^\infty}

\begin{document}

\title{Quantum metrology with one-dimensional superradiant photonic states}%
\date{\today}

\author{V. Paulisch}
\thanks{These authors contributed equally to this work.}
\affiliation{Max-Planck-Institut f\"ur Quantenoptik, Hans-Kopfermann-Str.~1, D-85748 Garching, Germany}
\author{M. Perarnau-Llobet}
\thanks{These authors contributed equally to this work.}
\affiliation{Max-Planck-Institut f\"ur Quantenoptik, Hans-Kopfermann-Str.~1, D-85748 Garching, Germany}
\author{A. Gonz\' alez-Tudela}
\email{a.gonzalez.tudela@csic.es}
\affiliation{Max-Planck-Institut f\"ur Quantenoptik, Hans-Kopfermann-Str.~1, D-85748 Garching, Germany}
\affiliation{Instituto de F\'isica Fundamental IFF-CSIC, Calle Serrano 113b, Madrid 28006, Spain.}
\author{J. I. Cirac}
\affiliation{Max-Planck-Institut f\"ur Quantenoptik, Hans-Kopfermann-Str.~1, D-85748 Garching, Germany}

\begin{abstract}
Photonic states with large and fixed photon numbers, such as Fock states, enable quantum-enhanced metrology but remain an experimentally elusive resource.
A potentially simple, deterministic and scalable way to generate these states consists of fully exciting $N$ quantum emitters equally coupled to a common photonic reservoir, which leads to a collective decay known as Dicke superradiance.
The emitted $N$-photon state turns out to be a highly entangled multimode state, and to characterise its metrological properties in this work we: (i) develop theoretical tools to compute the Quantum Fisher Information of general multimode photonic states; (ii) use it to show that Dicke superradiant photons in 1D waveguides achieve Heisenberg scaling, which can be saturated by a parity measurement; (iii) and study the robustness of these states to  experimental limitations in state-of-art atom-waveguide QED setups.
\end{abstract}
\maketitle

Quantum metrology exploits quantum resources, such as squeezing and entanglement, to enhance the  precision of measurements   beyond the capabilities of any classical scheme~\cite{Giovannetti2011,Tth2014,Dowling2015,demkowicz-dobrzanski15}. Given $N$ probes to estimate an unkown parameter $\varphi $, classical measurements are limited by the  shot-noise limit (SNL)  $\Delta \varphi = 1/\sqrt{N}$, whereas entangled probes can surprass this bound possibly reaching the Heisenberg limit (HL), $\Delta \varphi = 1/N$, which in fact provides the ultimate bound on sensitivity. In atomic ensembles, achieving quantum-enhanced metrology with relatively large particle numbers appears possible~\cite{Estve2008,Riedel2010,Ockeloen2013,LouchetChauvet2010,Appel2009,Fernholz2008,Wasilewski2010,Tth2010,Behbood2014}. The situation becomes more challenging when dealing with photonic states in optical interferometry. Squeezed states, a well known-resource~\cite{Caves1981},  are very challenging to scale up, with current demonstrations being at the few-photon level~\cite{Vahlbruch2016,Andersen2016}. 
States with a well-defined photonic number, e.g. NOON~\cite{Bollinger1996} and twin-Fock~\cite{holland93} states,  also constitute a powerful resource, which has been experimentally tested for few-photons states~\cite{mitchell04a,nagata07a,Slussarenko17}.
Yet, current experimental methods to generate these states  are limited by both low fidelities and efficiencies, since they are based in combining heralded single-photons with post-selection, which naturally leads to an exponential decrease of the efficiency with increasing $N$~\cite{dakna99a,wang16a}.
 
 A promising approach for generating multiphoton states in a deterministic, efficient and scalable manner are quantum emitters coupled to photonic waveguides~\cite{vetsch10aaa,thompson13a,goban13a,faez14a,  lodahl15a,sipahigil16a,corzo16a,sorensen16a,solano17a}.  In these setups, the waveguide decay rate, $\Gamma_\oned$, can exceed the free space one, $\Gamma^*$, and naturally enhance the photon collection efficiency of the system. On top of that, when all the quantum emitters couple equally to the waveguide, their dynamics is described by the celebrated Dicke model~\cite{Dicke1954aa}, which predicts an additional collective enhancement of the waveguide decay rate. Given $N$ emitters in the waveguide and $m$ collective atomic excitations, previous studies focused on the  regime  $m\ll N$~\cite{porras08aa}, where the $m$ collective excitations decay into a single-mode $m$-photon wavepacket with an error scaling as $\varepsilon_{\mathrm{lin}}\sim m\Gamma^*/(N\Gamma_\oned)$. The main limitation of 
this regime arises in the preparation of the initial state, since creating a fixed number $m$ of collective atomic excitations requires the use of sophisticated protocols~\cite{gonzaleztudela15a,paulisch17aa,gonzaleztudela17a}.

A conceptually and experimentally simpler approach consists of exciting all the quantum emitters, i.e., $m=N$. In this regime, the emitters experience a non-linear decay, known as Dicke superradiance, leading to a multimodal structure of the emitted $N$-photon wavepacket~\cite{gonzaleztudela15a}, which can be generally written as:
\begin{align}
\ket{\phi^{(N)}_A}
	=\multInt \frac{ \dd k_1 ... \dd k_N}{(2\pi)^N N!}    A_{ \{ k\}}  a^\dagger_{k_1}  \dots {a}^\dagger_{k_{N}} \ket{0},
\label{eq:multimode}
\end{align}
where $a_{k_i}^{\dagger}$ is the creation operator
of a waveguide photon of momentum $k_i$. The coefficient $A_{ \{ k \}} = A_{k_1, k_2, \cdots k_n}$ characterizes the multimodal structure of the wavepacket. 
In contrast to the case of linear decay processes~\cite{porras08aa}, it is not factorizable, $A_{ \{ k \}} \neq \sqrt{N!} A_{k_1} \cdots A_{k_n}$. 
This protocol uses all possible excitations while having a particularly simple initial state, making it very attractive for experiments. However, the multimode form of the emitted state prevents the direct use of previous results in quantum optical metrology~\cite{Giovannetti2011,Tth2014,Dowling2015,demkowicz-dobrzanski15}. In fact, the potential of Dicke superradiant states for metrology has not been addressed so far, despite being a promising candidate.

In this work, we show that one-dimensional Dicke superradiant states achieve Heisenberg scaling as $[\Delta\varphi]_{\rm Dicke} \approx 0.41 /N$, 
performing only slightly worse than Fock states, $[\Delta\varphi]_{\rm Fock} \approx 0.5/N$.  
Furthermore, we characterize the robustness of Dicke superradiant states to several experimental error sources, showing how they are particularly robust to photon losses 
 with an error scaling as $\varepsilon_{\mathrm{nl}}\propto\log(N) \Gamma^*/\Gamma_\oned$, for $\Gamma_\oned\gg \Gamma^*$. Thus, for a given ratio $\Gamma_\oned/\Gamma^*$, and desired error, $\varepsilon_{\mathrm{nl}}$, our protocol can potentially generate up to $N\sim \exp( \varepsilon_{\mathrm{nl}}\Gamma_\oned/\Gamma^*)$ photons, paving the way for efficient and scalable quantum-enhanced metrology protocols. To obtain these results, we develop theoretical tools to characterise the metrological properties of general multimode states of the form \eqref{eq:multimode} in Mach-Zender interferometry. We illustrate their potential  in multimodal photonic states created in anharmonic cavities~\cite{hartmann08a}, which we show to allow for quantum-enhanced metrology without reaching Heisenberg scaling,  and envisage they can be readily applied to other multimode states  that  appear in  relevant experimental setups, such as biexciton emission in quantum dots~\cite{ota11aa}.

\begin{figure}[t]
\includegraphics[width=0.47\textwidth]{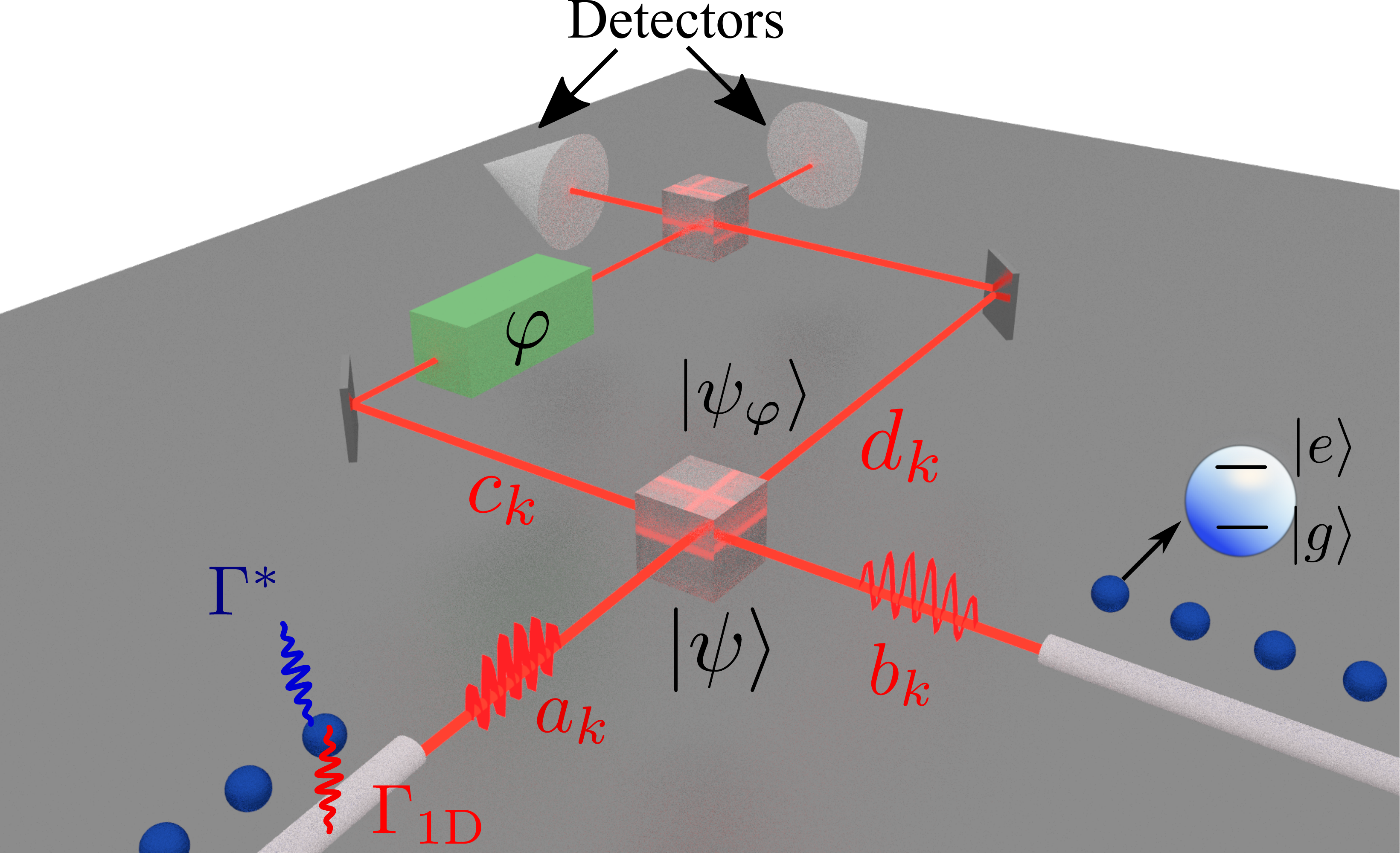}
\caption{General scheme of the protocol: two emitter ensembles are coupled collectively to two waveguides connected to the A/B ports of a Mach-Zehnder interferometer. 
}
\label{fig:MZI}
\end{figure}

\emph{Quantum optical interferometry.}
A paradigmatic task in optical interferometry is the  measurement of a phase $\varphi$ with high precision. The standard setup is the so-called Mach-Zehnder interferometer depicted in Figure \ref{fig:MZI}. The main resource of this protocol is the initial photonic state, $\ket{\psi}$, impinging onto the first beam splitter with input (output) ports $A/B$ ($C/D$) with annihilation operators $a/b$ ($c/d$). The beam splitter can be described  as a unitary $\bar{U}_{BS}(\theta)=\exp[\theta (a^{\dagger}b-b^{\dagger}a)]$ with the mixing angle $\theta$. After this operation the photon can travel in two different arms, acquiring a relative phase $\varphi$ through $\bar{U}_\varphi=\exp(-i \varphi/2 (c^{\dagger}c-d^{\dagger}d))$.
This results in a state $\ket{\psi_{\varphi}}$, which now contains  information on $\varphi$. By applying a measurement $M$ on $\ket{\psi_{\varphi}}$, the phase $\varphi$ can be estimated with an uncertainty $\Delta \varphi$.
In general, $\Delta \varphi$ depends on $\ket{\psi_{\varphi}}$,  $M$, and the number of repetitions of the experiment $\nu$. Assuming $\nu\gg 1$, the Quantum Cram\'er-Rao Bound~\cite{helstrom76,holevo82} gives a lower bound for $\Delta \varphi$ that is independent of $M$, $(\Delta \varphi)^2 \geq 1/\nu F_Q [\psi_\varphi ],$
%
%
where  $F_Q [\psi_\varphi]$ is the Quantum Fisher Information (QFI)~\cite{Braunstein1994} of the state $\ket{\psi_\varphi}$, $F_Q [\psi_\varphi]
	= 4 \left( \langle \dot{\psi}_\varphi | \dot{\psi}_\varphi \rangle 
			- |\langle \dot{\psi}_\varphi | \psi_\varphi \rangle |^2
			\right)$.
%
The QFI characterises the  potential of $\ket{\psi_{\varphi}}$ for estimating $\varphi$ with an optimal measurement. 

Let us illustrate the power of the QFI, with relevant examples in optical interferometry. For example, in classical sources using coherent states, $\ket{\psi^{\mathrm{cl}}}=\ket{\alpha}_A\otimes\ket{0}_B$, the QFI is at most proportional to the average photon number, $F_Q\le \bar{N}$, with $\bar{N}=|\alpha|^2$.
The upper bound of the QFI, given by the HL, $F_Q=N^2$, is obtained by  NOON states~\cite{Bollinger1996},  $\ket{\psi_{\varphi}^\mathrm{NOON}}=\frac{1}{\sqrt{2}}\left(\ket{N 0}+\ket{0 N}e^{i N\varphi}\right)$, where $\ket{N_u N_l}$ indicates the number of photons in the C/D path. The more experimentally friendly Twin-Fock States (TFS)~\cite{holland93},  $\ket{\psi^{\rm TFS}_{\varphi}}= \bar{U}_{\varphi}\bar{U}_{BS}(\pi/4)\ket{N/2}_A\otimes\ket{N/2}_B$, obtained when two  Fock states enter into the first beam splitter, also lead to Heisenberg scaling with slightly worse slope,
\begin{align}\label{eq:QFI_TFS}
F_Q[\psi^{\rm TFS}_{\varphi}] = \frac{N(N+2)}{2}\,.
\end{align}
Furthermore, this bound can be saturated by  a number-resolved measurement~\cite{Pezze2013} or a parity~\cite{campos03a} measurement, which is optimal for any bosonic state that is mode-symmetric~\cite{Hofmann2009}. Many other two-mode quantum states enable quantum-enhanced metrology (see e.g.  \cite{Olivares2007,Pezz2009,Hyllus2010}), notably including  random bosonic states \cite{Oszmaniec2016}. In the following section, we go beyond the standard two-mode interferometry described above (see e.g. the review \cite{demkowicz-dobrzanski15}), and analyse the QFI when the input states of the interferometer are   multimode states of the general form~\eqref{eq:multimode}.

\emph{Quantum Fisher information of multimode states.}
Let us restrict our attention to the case where the initial state $\ket{\psi}=\ket{\phi_A^{N_A}}\otimes \ket{\phi_B^{N_{B}}}$ has a well defined photon number $N_{A/B}$ at the A/B ports of the first beam splitter.  The total photon number $N=N_A+N_B$ is the metrological resource. The states have the multimodal structure  \eqref{eq:multimode} with modal coefficients $A_{\{k\}}/B_{\{q\}}$, and where $\{k\}=\{k_1,\dots,k_{N_A}\}$ and $\{q\}=\{q_1,\dots,q_{N_B}\}$ represent the internal degrees of freedom of the A/B wavepacket. In our case  they are the momenta of the  photons in the A/B wavepackets, although the problem is generally formulated.
 
Generalizing the beam splitter and phase operation to deal with multimode variables: $U_{\rm BS}(\theta)=\exp[\int\frac{\dd k}{2\pi}(a_k^{\dagger}b_k-b_k^{\dagger}a_k)\theta]$ and $U_\varphi = \exp\left[-\ii \frac{\varphi}{2} \int \frac{\dd k}{2\pi} (c_k^\dagger c_k-d_k^\dagger d_k)\right]$, we consider states of the form $\ket{\psi^{\rm AB}_{\varphi}}=U_{\varphi}U_{\rm BS}(\pi/4))\ket{\phi^{(N_A)}_A}\otimes \ket{\phi^{(N_B)}_B}$. 
Exploiting the bosonic symmetry of the wavepackets $A_{\{k\}}/B_{\{k\}}$ under permutation, we simplify the QFI of $\ket{\psi^{\rm AB}_{\varphi}}$ to a very transparent formula (see Supplementary Material Sec.~I), $F_Q \left[\psi^{\rm AB}_{\varphi} \right] = 2 N_A N_B I_{AB} + N_A +N_B.$
%
%
which only depends on a single integral $I_{AB}$:
\begin{align}
I_{AB}=& \multInt \frac{\prod_{i,j=1,1}^{N_A,N_B}\dd k_i \dd q_j}{(2\pi)^{N_A+N_B} N_A! N_B!} 
		A^*_{k_1,...,k_{N_A}}B^*_{q_1,...,q_{N_B}}\times \nonumber\\
		& A_{q_1,k_2,...,k_{N_A}}B_{k_1,q_2...,q_{N_B}},\label{eq:genint}
\end{align} 
where the two indices $k_1/q_1$ have been exchanged in one of the coefficients. This formula is applicable to general multimode photonic states of a fixed photon number, and in the Supplementary Material Sec.~I we extend it to situations where the number of photons is only fixed in one input of the interferometer. It is easy to see that in the single-mode case $I_{AB}=1$, in agreement with  previous results \cite{Pezze2013}. 
Let us  now focus on the case where the A/B wavepackets have the same number of photons $N_A=N_B=N/2$ and the same modal structure $A_{\{k\}}=B_{\{k\}}$. These twin multimode states (TMS), denoted as $\ket{\psi^{\rm TMS}_{\varphi}}$, have a simple expression for the QFI,
\begin{align}
F_Q[\psi^{\rm TMS}_{\varphi}] = \frac{N (I_N N+2)}{2}\,,
\label{eq:FTMS}
\end{align}
where $I_N$ is the integral of Eq.~\eqref{eq:genint} for $N_A=N_B=N/2$ and $A_{\{k\}}=B_{\{k\}}$. Thus, a general multimode wavepacket will beat the SNL as long as $I_N$ decays slower than $1/N$, and reach HL scaling if $I_N$ tends to a constant. 
Importantly, in Sec.~II of the Supplementary Material  we  show that the QFI saturates for a parity measurement. Now  we  compute $I_N$ of two experimentally relevant  photonic states, Dicke superradiant states and photonic states generated in anharmonic cavities. 

\emph{QFI of one-dimensional superradiant states}.
The first multimode photonic states that we consider are the ones naturally generated from $N$ fully excited quantum emitters, described as two-level systems $\{\ket{g},\ket{e}\}$, with an optical transition coupled to a waveguide mode at a rate $\Gamma_\oned$. We focus on the so-called mirror configuration~\cite{corzo16a,sorensen16a,solano17a}, in which the emitter positions are fixed such that all of them interact equally with the waveguide modes. In that configuration, and assuming that the relaxation time-scales of the waveguide are much faster than the time-scales of the system dynamics~\cite{gardiner_book00a}, the quantum emitter dynamics are governed by the Dicke model~\cite{Dicke1954aa} $ \dot{\rho}=i(\rho H^\dagger_\mathrm{eff} - H_\mathrm{eff}\rho)+\Gamma_\oned S_{ge}\rho S_{eg}$, where $\rho$ is the density matrix describing the quantum emitters' state, $H_\mathrm{eff} = \omega_0  S_{ee} - \ii \frac{\Gamma_\oned}{2} S_{eg} S_{ge}$ the effective non-Hermitian Hamiltonian, and where we denote the collective emitter operators as $S_{\alpha \beta} = \sum_j \ket{\alpha}_j \bra{\beta}$.

Interestingly, if we initialize the system to be fully excited, $\ket{e}^{\otimes N}$, both the effective Hamiltonian and the quantum jump terms $S_{ge},S_{eg}$  restrict the evolution to the fully symmetric space. This guarantees that only $N$ states participate in the evolution, which can be classified depending on their number of excitations, $m$, that we denote as $\ket{\psi_m} \propto S_{eg}^m \ket{g}^{\otimes N}$. From $H_\mathrm{eff}$, note that these energy levels are linearly spaced, $\omega_m=m\omega_0$. The non-linearity of the process emerges from the  decay rates $\gamma_m$  (associated to the transition $\ket{\psi_m} \rightarrow \ket{\psi_{m-1}}$) which depend on the number of excitations as  $\gamma_m=m(N-m+1)\Gamma_\oned$. In particular, $\ket{\psi_m}$ decays as $\gamma_m\propto N$ at the beginning and end of the process, while accelerating in the middle part, where $\gamma_{N/2}\propto N^2$, refered to 
as Dicke superradiance effect. When all the emitters have decayed, the resulting photonic state reads~\cite{gonzaleztudela15a}
\begin{align}
\hspace{-.15cm}
A_{\{k \}}\hspace{-.05cm}
	= \hspace{-.05cm}\prod_{j=1}^N \frac{\sqrt{\gamma_j}}{\ii(j \omega_0 - \sum_{i=1}^j \omega_{k_i}) - 
	 \frac{1}{2}\gamma_j} + \big\{ k_i \leftrightarrow k_j \big\},\hspace{-.05cm}
\label{eq:Ak}
\end{align} 
where we use the notation $ \big\{ k_i \leftrightarrow k_j \big\}$ to denote that the expression has to be symmetrized with respect to the momenta $k_j$. This wavepacket inherits the non-linearity from the decay process as temporal correlations between the $N$ photons, and thus, it can not be factorized as a single-mode one.

\begin{figure}[tb]
\includegraphics[width=0.47\textwidth]{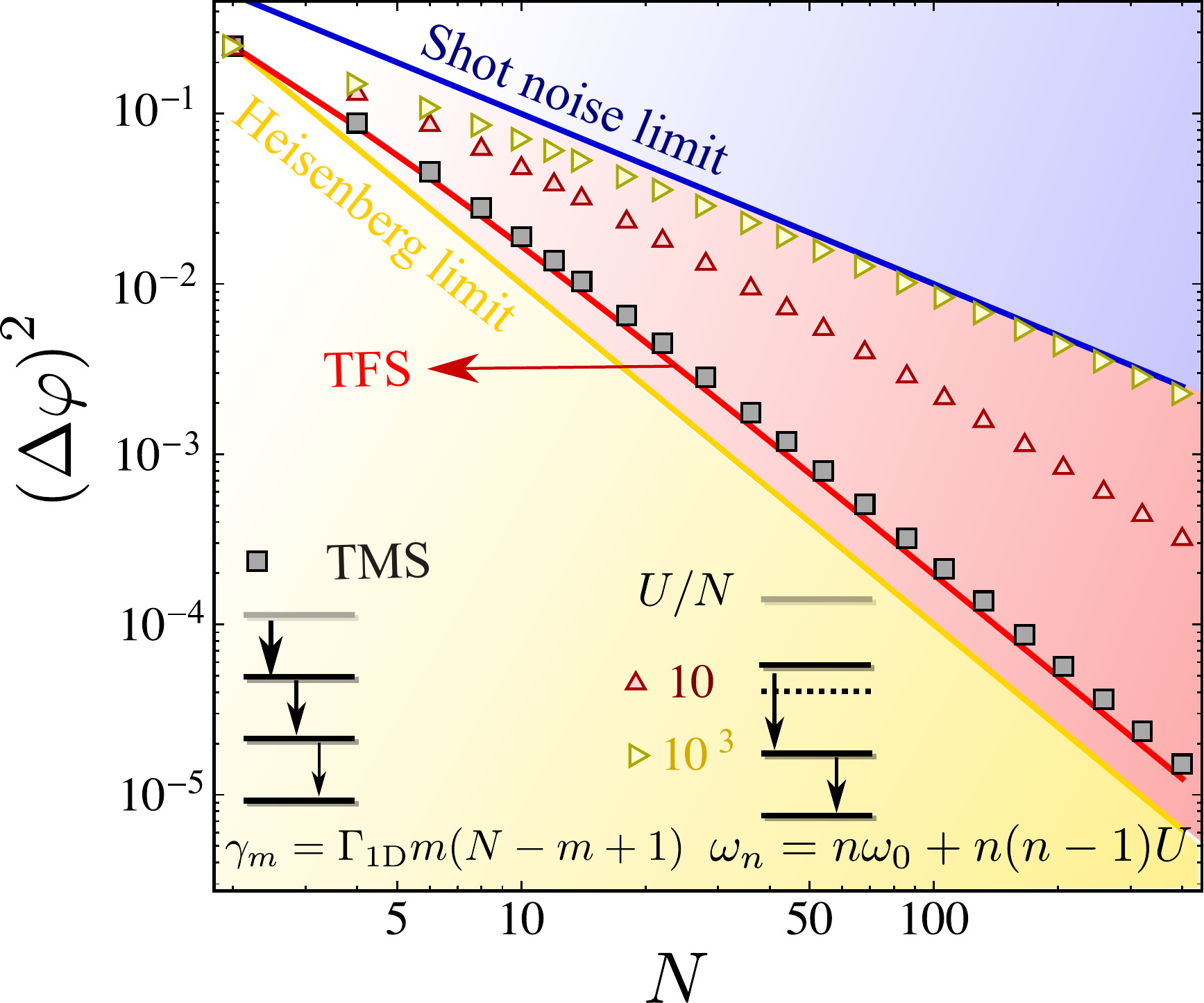}
\caption{Scaling of $(\Delta\varphi)^2$ with $N$ for several situations discussed along the manuscript. In solid blue/yellow we plot both the shot-noise and Heisenberg limit, respective. In solid red, the scaling of Fock states. In black squares, we plot the numerical results for TMDS. In red/yellow triangles. we plot the numerical results obtained of anharmonic cavities for $U/N=10/10^3$, respectively.}
\label{fig:Phase}
\end{figure}

We now study the metrological potential of the states \eqref{eq:Ak}. For that,
we consider a protocol where two ensembles with $N/2$ emitters are placed at the input ports of a Mach-Zehnder interferometer, as depicted in Fig.~\ref{fig:MZI}. Then,  a simultaneous collective $\pi$-pulse is implemented in both ensembles, leading to an emission of  two heralded $N/2$-photon states: $\ket{\phi_A^{(N/2)}}$, $\ket{\phi_B^{(N/2)}}$~\cite{Note3}. For obtaining the QFI of this state, we  need to calculate the $N$-variable integral $I_N$ for $A_{\{k \}}$ defined in Eq.~\eqref{eq:Ak}. Note that there are $(N/2)!$ terms in~\eqref{eq:Ak}, which lead to $((N/2)!)^4$ terms in the integral. The number of integrals can always be reduced to $(N/2)^4$  by noting that all variables in $I_N$ are exchangeable  except for $q_1$ and $k_1$~\cite{Note2}. For the  $A_{\{k \}}$ in~\eqref{eq:Ak},  we develop a recurrence relation 
which can be efficiently computed for large $N$, consisting of a multiplication of $N$ matrices of 
at most size $3N \times 3N$  (see Sec.~IV of the Sup.~Material). This allows for determining $I_N$ exactly for large $N$.   With this method, we numerically obtain that $I_N$ quickly approaches a constant $I_N\approx 0.82$ for the range of $N$ considered (up to approx. $500$ photons). This has the important consequence that the QFI of superradiant TMDS shows the Heisenberg scaling:
\begin{align}
F_Q[\psi^{\rm TMDS}_{\varphi}] \approx 0.41N^2+N.
\label{QFITMS}
\end{align} 
where TMDS stands for twin multimode Dicke states.
In Fig.~\ref{fig:Phase}, we plot $(\Delta\varphi)^2$ of $\psi^{\rm TMDS}_{\varphi}$ in black squares, together with Fock states (in solid red) showing how the multimodal case has the same scaling, just with a slightly reduced prefactor. This is the most important result of this work, since it provides a path towards efficient and scalable multiphoton states useful for quantum metrology protocols.

\emph{Experimental considerations.}
Since there are currently many platforms~\cite{vetsch10aaa,thompson13a,goban13a,faez14a,
  lodahl15a,sipahigil16a,corzo16a,sorensen16a,solano17a} with the potential to obtain superradiant photonic states, we analyze now the resilience of the QFI to several experimental imperfections in the preparation stage. We start by considering the main source of noise of these setups which comes from the emission into  modes other than the waveguide ones, e.g., free-space or a different waveguide polarization, that we embed into a single (individual) decay rate $\Gamma^*$.  This term takes the states $\ket{\psi_m}$ out of the collective subspace at a rate $m \Gamma^*$, so that it is especially critical when the system is fully excited. The probability of emitting $N$-collective photons, 
which translates into a photon state fidelity, can be estimated as the probability of no-jump in each step:
\begin{equation}
p\approx \prod_{m=1}^{N} \left(1-\frac{m \Gamma^*}{\gamma_m \Gamma_{\oned}}\right) \approx 1-\ln(N)\frac{\Gamma^*}{\Gamma_\oned}
\label{ploss}
\end{equation}
which is a valid assumption as long as $\Gamma_\oned\gg \Gamma^* \ln(N)$ and $N\gg 1$, like we numerically confirm through exact integration of the master equation (see Sec.~V of the Sup. Material). The resulting photonic state will be a mixed state which can be written as: $\rho_{N,*}= p \ket{\phi_A^{(N)}} \bra{\phi_A^{(N)}} + (1-p) \sigma_1$ where  $\sigma_1$ is a convex combination of state with less than $N$ photons in the waveguide. Using two such mixed states as input of the interferometer, $\rho^{\rm TMDS}=\rho_{N/2,*}\otimes \rho_{N/2,*}$, we can bound its QFI by noting that the 
 the QFI is non-negative and additive under direct sum, obtaining $F_Q[\rho^{\rm TMDS}_{\varphi}]\geq p^2  F_Q[\psi^{\rm TMDS}_{\varphi}].$
%
 This shows that, as long as we are in the limit  $\Gamma_\oned\gg \Gamma^*$, 
the results become robust to photon loss with an error that increases only logarithmically with $N$.   

Let us now enumerate other error sources, and provide the conditions under which they can be neglected (see Sec.~V of the Sup. Material for details). Absorption within the waveguide (or scattering through imperfections) provides a finite propagation length to the waveguide modes, $L_\mathrm{prop}$, which spoils the collective behaviour of the atomic interactions. To be able to neglect this effect the propagation length must be larger than the system size $L_\mathrm{prop}\gg N \lambda_a$, being $\lambda_a$ the wavelength of the waveguide modes determining the distance between QEs.  State-of-art values for SiN waveguides~\cite{goban13a} show $L_\mathrm{prop}/\lambda_a\sim 5\times 10^4 $, such that this will be in general a small correction. This finite lifetime of the waveguide modes also leads to photon loss while the wavepacket propagates away from the QEs.  
Furthermore, to neglect retardation effects, 
the propagation timescales, $\sim N\lambda_a/v_g$ must be much shorter than the shortest emitter timescale, that in this case occurs in the middle of the superradiant decay, being proportional to $(\Gamma_\oned N^2/4)^{-1}$. 

Another error sources is the deviation from the initial atomic state, e.g., by an imperfect control of the timing, $T$, or laser amplitude, $\Omega$, in $\pi$-pulse, that we embed in a single parameter $\Delta (\Omega T)$. If $\Delta (\Omega T) \sqrt{N}\ll 1,$ this translates into a different initial state, $\approx \left(1-i \Delta (\Omega T)\sqrt{N} S_{ge}\right)\ket{e}^{\otimes N}$, which leads to an error scaling as $\sim \Delta (\Omega T)^2 N$. Other deviations from the ideal setting are that the two QE ensembles couple differently to the waveguide, $\Delta\Gamma_\oned=\Gamma_\oned-\Gamma_\oned^{'}$, or that the wavepackets are emitted with a certain time delay, $\tau$, rather than simultaneously. Both deviations decrease the integral $I_N$ in the following way: $I_{N,\Delta\Gamma_\oned}\approx I_N \left(1 - \frac{(\Delta\Gamma_\oned)^2}{8\Gamma_\oned^2} N\right)$ and $I_{N,\tau}\gtrsim I_N \left( 1 - N \Gamma_\oned \tau \right)$ for $\Delta\Gamma_\oned\ll \Gamma_\oned$ and $N\Gamma_\oned\tau\ll 1$, 
respectively. Summarising, one must ensure simultaneously that $N(\Delta\Gamma_\oned)^2 / \Gamma_\oned^2 \ll 1 $ and $N\Gamma_\oned\tau\ll 1$ in order to guarantee quantum-enhanced metrology. 

Finally, let us now briefly  discuss imperfections in the interferometer and in the measurement.  In Sec. VF of the Sup. Material, we find that photon loss in the interferometer, quantified by a probability $\eta$, leads to a correction to the QFI that is smaller than $\eta N^2 I_N /4$ (at first order in $\eta$), so that one needs at most $\eta \ll 4.9 N^{-2}$ to ensure Heisenberg scaling. For sufficiently large $N$, Heisenberg scaling is eventually lost with photon loss independently of the state into consideration. Then, the quantum advantage just shows up as a better prefactor in the scaling of the QFI with $N$ \cite{escher2011general,janekdecoherence}. 
This regime, which requires dealing with mixed states, will be discussed in a forthcoming publication~\cite{Perarnau2019}, where we will show that TMDS of $N$ photons have similar metrological properties than twin-Fock states of $\approx 0.91N$ photons. 
Given that twin-Fock states are known to be robust to photon loss, both in the interferometer \cite{Knysh2011} and in the measurement device~\cite{Pezze2013},  we expect  TMDS to be a valuable resource for quantum-enhanced metrology in the presence of photon loss in the interferometer and in the apparatus~\cite{Perarnau2019}.

\emph{QFI of anaharmonic cavities.}
Let us  illustrate the potential of the tools we developed with another class of non-linear photonic states appearing from anharmonic cavities~\cite{hartmann08a}, where the non-linearity manifests as an non-linear energy shift, i.e., $\omega_n=n\omega_0+n(n-1) U$, while having linear decay rates $\gamma_n=n\gamma_1$. In Fig.~\ref{fig:Phase} we plot the $(\Delta\varphi)^2$ for the photonic state emerging from the decay at $N$-th level of the anaharmonic ladder for two values of $U/\gamma_1$. Interestingly, we observe  that Heinseberg scaling is lost for any value of $U$, as we find numerically that $I_N \propto 1/N$ for large enough $N$. This is illustrated in Figure \ref{fig:Phase} for $U/\gamma_1=10,10^3$. This result shows that different multimode states can have completely different metrological properties, suggesting a rich relation between the multimode structure and the potential for metrology of the state. It also provides  intuition on why TMDS behave similarly than TFS, since in that 
case all photons are spectrally centred at the same frequency $\omega_0$, thus being mostly indistinguishable.

\emph{Conclusions.}
To sum up, we have proven that photons emitted from Dicke superradiant states~\cite{Dicke1954aa} are useful for quantum metrology. To show it, we derive a computationally friendly way of calculating the QFI for arbitrary multimode photonic wavepackets, illustrating its power with another physically relevant example (photons emitted from anharmonic cavities). The number of photons that can be produced for a fixed fidelity scales exponentially with $\Gamma_\oned/\Gamma^*$. State-of-art nanophotonic setups~\cite{lodahl15a} have already achieved ratios $\approx 60$ with  $\Gamma_\oned\sim 1$ GHz, which indicates the possibility of generating hundred of photons at the level of 90\% fidelities and GHz rates. Furthermore, most of the conclusions can be extrapolated to other systems where collective decays can be engineered, such as cavity QED setups~\cite{haas14a,norcia16a,hosseini17a,kim17a}.
 We foresee other possible applications in situations where Fock states provide advantage, as it is the case in quantum lithography~\cite{boto00a}.

\emph{Acknowledgements.} We thank  J. B. Brask, C. Gogolin, P. Hofer,  J. Ko\l{}ody\'{n}ski, V. Sandoghdar and J. Thompson for valuable discussions and comments on the manuscript.
VP  acknowledges  the  Cluster  of  Excellence  Nano Initiative Munich (NIM). MPL acknowledges the Alexander Von Humboldt Foundation.  AGT and IC acknowledge ERC Advanced Grant  QENOCOBA under the EU Horizon 2020 program (grant agreement 742102).

\bibliography{ref_vip,Sci,books,Sci-v2}

\newpage

\setcounter{equation}{0}
\setcounter{figure}{0}
\setcounter{section}{0}
\makeatletter

\renewcommand{\thefigure}{SM\arabic{figure}}
\renewcommand{\thesection}{SM\arabic{section}}  
\renewcommand{\theequation}{SM\arabic{equation}}

\begin{widetext}
	
		\textbf{\large Supplementary Material: Quantum metrology with one-dimensional superradiant photonic states}

\section{Derivation of the Quantum Fisher Information}
\label{App.Sec.QFI}

To derive the Quantum Fisher Information of the TMS, we work in a more general setting of two different multi-mode states incident on the two ports of the first beamsplitter. We write them as
\begin{align*}
\ket{\phi^{(m)}_A}
&=\multInt \frac{ \dd k_1 ... \dd k_m}{(2\pi)^m m!} 
A_{k_1 \ldots k_m}  a^\dagger_{k_1}  \dots {a}^\dagger_{k_m} \ket{0},\\
\ket{\phi^{(n)}_B}
&=\multInt \frac{ \dd \tilde{k}_1 ... \dd \tilde{k}_n}{(2\pi)^n n!} 
B_{\tilde{k}_1 \ldots \tilde{k}_n}  b^\dagger_{\tilde{k}_1}  \dots b^\dagger_{\tilde{k}_n} \ket{0}.
\end{align*}
Abusing notation, here we will also call the output modes of the interferometer by $a_k$ and $b_k$. Then, the state after the beam splitter and the phase operation, $\ket{\psi_\varphi} = U_{\varphi} U_\mathrm{BS}  \ket{\phi^{(m)}_A}\otimes \ket{\phi^{(n)}_B}$, is  given by
\begin{equation}
\ket{\psi_\varphi}
= \multInt \frac{\prod_{i,j=1}^{m,n} \dd k_i \dd \tilde{k}_j }{(2\pi)^{m+n} m! n!}
A_{k_1 \ldots k_m} B_{\tilde{k}_1 \ldots \tilde{k}_n} 
c^\dagger_{k_1}  \dots c^\dagger_{k_m}   d^\dagger_{\tilde{k}_1} \dots d^\dagger_{\tilde{k}_n} \ket{0}.
\end{equation}
where we defined the creation operators  $c_k^{\dagger} \equiv c_k^\dagger (\varphi) = U_{\varphi} U_\mathrm{BS} a_k^{\dagger} U_\mathrm{BS}^\dagger U_\varphi^\dagger =  \frac{1}{\sqrt{2}}\left(e^{-\ii \varphi/2}a_k^\dagger + \eexp{\ii \varphi/2} b_k^\dagger \right)$, and respectively for $d_k^{\dagger} \equiv d_k^\dagger (\varphi) = \frac{1}{\sqrt{2}} \left(-e^{-\ii \varphi/2}a_k^\dagger + \eexp{\ii \varphi/2} b_k^\dagger \right)$.

For calculating the QFI of $\ket{\psi_\varphi}$ one has to take the derivative, which turns  $c_k^{\dagger}$ ($d_k^{\dagger}$)  into $\frac{i}{2}d_k^{\dagger}$  ($\frac{i}{2}c_k^{\dagger}$). This leads to
\begin{equation}
\ket{\dot{\psi}_\varphi}
= \left(\frac{i}{2} \right)^{\frac{m+n}{2}}\multInt \frac{\prod_{i,j=1}^{m,n} \dd k_i \dd \tilde{k}_j }{(2\pi)^{m+n} m! n!}
A_{k_1 \ldots k_m} B_{\tilde{k}_1 \ldots \tilde{k}_n} 
\left( m d^{\dagger}_{k_1}d^{\dagger}_{\tilde{k}_1}+nc^{\dagger}_{k_1} c^{\dagger}_{\tilde{k}_1} \right)
c^\dagger_{k_2}  \dots c^\dagger_{k_m}   d^\dagger_{\tilde{k}_2 } \dots d^\dagger_{\tilde{k}_n} \ket{0}.
\label{psi_varphi}
\end{equation}
where we used the symmetry of $A_{k_1 \cdots k_m}$ and $B_{k_1 \cdots k_n}$ under permutations of $k$'s. The expressions $\left\vert \braket{\psi_\varphi}{\dot{\psi}_\varphi}  \right\vert$ and $\braket{\dot{\psi}_\varphi}{\dot{\psi}_\varphi}$, which determine the QFI, can be evaluated by using the commutation relations 
$\left[c_p, d_k^\dagger \right] = \left[d_p, c_k^\dagger \right] = 0$,
$\left[c_p, c_k^\dagger \right] = \left[d_p, d_k^\dagger \right] = 2 \pi \delta(p - k)$. It is clear that,
\begin{align}
\left\vert \braket{\psi_\varphi}{\dot{\psi}_\varphi}  \right\vert=0
\end{align}
as  $\ket{\psi_\varphi}$ and  $\ket{\dot{\psi}_\varphi}$  contain a different number of $c$'s and $d$'s. To compute  $\braket{\dot{\psi}_\varphi}{\dot{\psi}_\varphi}$, we use the symmetry of $A_{k_1 \cdots k_m}$ and $B_{k_1 \cdots k_n}$, which allows us  to take one representative of each of the $c_k$'s or $d_{\tilde{k}}$'s and multiply by the number of times it appears. One has to evaluate an integral over correlation functions $f(X)$ over the $k$'s and $\tilde{k}$'s, for which we introduce the shorthand notation
\begin{equation}
\int \dd X f(X) 
\equiv \multInt \frac{\prod_{i,j}^{m,n} \dd k_i \dd \tilde{k}_j \dd p_i \dd \tilde{p}_j}{(2\pi)^{2(m+n)} m!^2 n!^2} 
A_{p_1 \cdots p_m}^* B_{\tilde{p}_1 \cdots \tilde{p}_n}^* A_{k_1 \cdots k_m} B_{\tilde{k}_1 \cdots \tilde{k}_n} f(X)
\end{equation}
Then we can write
\begin{align}
\braket{\dot{\psi}_\varphi}{\dot{\psi}_\varphi} 
&=
\frac{1}{4} \int \dd X\
\bra{0} d_{\tilde{p}_n} \dots d_{\tilde{p}_2} c_{p_m} \dots c_{p_2} 
\left(   m d_{p_1}d_{\tilde{p}_1}+nc_{\tilde{p}_1}c_{\tilde{p}_1} \right)
\left( m d^{\dagger}_{k_1}d^{\dagger}_{\tilde{k}_1}+nc^{\dagger}_{k_1} c^{\dagger}_{\tilde{k}_1} \right)
c^\dagger_{k_2}  \dots c^\dagger_{k_m}   d^\dagger_{\tilde{k}_2 } \dots d^\dagger_{\tilde{k}_n} \ket{0} 
\notag \\
&=\frac{1}{4}\int \dd X
m^2 \Bigg(
\left[d_{p_1}, d^\dagger_{k_1} \right] (m-1) \left[c_{p_2}, c^\dagger_{k_2} \right] n \left[d_{\tilde{p}_1}, d^\dagger_{\tilde{k}_1}\right]+
n\left[d_{p_1}, d^\dagger_{\tilde{k}_1} \right] n \left[d_{\tilde{p}_1}, d^\dagger_{k_1} \right] (m-1) \left[c_{p_2}, c^\dagger_{k_2}\right]\Bigg)
\notag \\
& \qquad \qquad		
\qquad \qquad
\times (m-2)! \prod_{i=3}^m \left[c_{p_i}, c^\dagger_{k_i}\right]
(n-1)! \prod_{j=2}^n \left[d_{\tilde{p}_j}, d^\dagger_{\tilde{k}_j}\right]
+ m \leftrightarrow n 
\notag \\
&=
\frac{1}{4}\int \dd X\ m! n! \prod_{i=2}^m 2 \pi \delta(p_i - k_i)
\prod_{j=2}^n 2 \pi \delta(\tilde{p}_j - \tilde{k}_j)
\notag \\
& \qquad \qquad \qquad
\times \left[ (m+n)
(2\pi)^2 \delta(p_1 - k_1) \delta(\tilde{p}_1 - \tilde{k}_1) 
+ 2nm
(2\pi)^2 \delta(p_1 - \tilde{k}_1) \delta(\tilde{p}_1 - k_1) 
\right]
\notag \\ 
&
=\frac{1}{4} \left( (m+n) I_{AB}^\klamm{0} 
+ 2 m n I_{AB}^\klamm{1} \right).
\end{align}
where the $I_{AB}^\klamm{0}$ and $I_{AB}^\klamm{1}$ read:
\begin{subequations} \label{eq:Ims}
	\begin{align}
	I_{AB}^\klamm{0}
	=& \multInt \frac{\prod_{i,j=1,1}^{m,n}\dd k_i \dd \tilde{k}_j}{(2\pi)^{m+n} m! n!} 
	\left\vert A_{k_1,...,k_m} \right\vert^2 
	\left\vert B_{\tilde{k}_1,...,\tilde{k}_m} \right\vert^2
	= 1,
	\\
	I_{AB}^\klamm{1}
	=& \multInt \frac{\prod_{i,j=1,1}^{m,n}\dd k_i \dd \tilde{k}_j}{(2\pi)^{m+n} m! n!} 
	A^*_{k_1,...,k_m}B^*_{\tilde{k}_1,...,\tilde{k}_n} 
	A_{\tilde{k}_1,k_2,...,k_m}B_{k_1,\tilde{k}_2...,\tilde{k}_n},
	\end{align}
\end{subequations}
where in $I_{AB}^{(1)}$  the two indices $k_1$ and $\tilde{k}_1$ have been exchanged in one of the coefficients. The QFI is then given by:
\begin{equation*}
F_Q \left[\psi_\varphi \right] 
= 4 \left(\braket{\dot{\psi}_\varphi}{\dot{\psi}_\varphi} 
- \left\vert \braket{\psi_\varphi}{\dot{\psi}_\varphi}  \right\vert^2 \right)
= 2 m n I_{AB}^\klamm{1} + m+n.
\end{equation*}
This provides the desired result. 

Finally, we note that this result can be easily generalised to states of the form
\begin{align*}
\ket{\phi^{(m)}_A}
&=\multInt \frac{ \dd k_1 ... \dd k_m}{(2\pi)^m m!} 
A_{k_1 \ldots k_m}  a^\dagger_{k_1}  \dots {a}^\dagger_{k_m} \ket{0},\\
\ket{\phi_B}
&=\sum_n  c_n \multInt \frac{ \dd \tilde{k}_1 ... \dd \tilde{k}_n}{(2\pi)^n n!} 
B^{(n)}_{\tilde{k}_1 \ldots \tilde{k}_n}  b^\dagger_{\tilde{k}_1}  \dots b^\dagger_{\tilde{k}_n} \ket{0}.
\end{align*}
That is, when an arbitrary pure   with $m$ photons enters one arm, and an arbitrary pure state enters into the other. By noting that states with a different total photon number do not mix, we can use  our previous derivation to arrive at the following QFI of  $\ket{\psi_\varphi'} = U_{\varphi} U_\mathrm{BS}  \ket{\phi_A^{(m)}}\otimes \ket{\phi_B}$,
\begin{align}
F_Q \left[\psi_\varphi' \right] = 2m \sum_n \left( |c_n|^2  n  I_{AB}^{\klamm{1},n} \right) + m + \langle n \rangle 
\end{align}
with $ \langle n \rangle = \sum_n  |c_n|^2n$, and 
\begin{align}
I_{AB}^{\klamm{1},n}  =& \multInt \frac{\prod_{i,j=1,1}^{m,n}\dd k_i \dd \tilde{k}_j}{(2\pi)^{m+n} m! n!} 
A^*_{k_1,...,k_m}B^{(n)*}_{\tilde{k}_1,...,\tilde{k}_n} 
A_{\tilde{k}_1,k_2,...,k_m}B^{(n)}_{k_1,\tilde{k}_2...,\tilde{k}_n}.
\end{align} 
This result extends one of the results of Ref. \cite{Pezze2013} on Fock states to arbitrary  photonic  states of a fixed photon number. We also note that this result holds for pure states, leaving the extension to mixed states as an interesting challenge for the future. 

\section{Parity Measurement}
\label{App.Sec.PM}

We now show that a parity measurement after the second beamsplitter transformation of the MZI locally resolves the phase at the Heisenberg limit when $n=m=N/2$, which is the case of main interest. Strictly speaking, this is achieved in the limit $\varphi\rightarrow 0$, but one can always add phase shifters during the estimation processing so that this does not rest generality~\cite{Giovannetti2011,demkowicz-dobrzanski15}.

The measurement operator can be written as $O = U_\mathrm{BS}^\dagger (-1)^{\int \frac{\dd k}{2 \pi }  a_k^\dagger a_k} U_\mathrm{BS}$, where the beamsplitter transformation is generated by $U_\mathrm{BS}=\mathrm{exp}\left[\int \frac{\dd k}{2 \pi } \ii (a_k^\dagger b_k - b_k^\dagger a_k) \pi /4\right]$, such that 
\begin{equation}
O 
= \prod_k 
\eexp{-(a_k^\dagger b_k - b_k^\dagger a_k) \pi /4} 
\eexp{\ii \pi a_k^\dagger a_k}
\eexp{(a_k^\dagger b_k - b_k^\dagger a_k) \pi /4}
= \prod_k
\eexp{\ii (a_k^\dagger - b_k^\dagger) (a_k - b_k) \pi/2}.
\end{equation}
We used the transformation $U_\mathrm{BS}^\dagger a_k^\dagger U_\mathrm{BS} = \frac{1}{\sqrt{2}} \left(a_k^\dagger - b_k^\dagger \right)$. Because $O^2 = 1$, the phase variance around $\varphi \approx 0$ is
\begin{equation}
\Delta \varphi^2 
= \lim_{\varphi \rightarrow 0} \frac{\langle \Delta O^2 \rangle}{(\partial_\varphi \langle O \rangle)^2}
= \lim_{\varphi \rightarrow 0} \frac{1- \langle O \rangle^2}{(\partial_\varphi \langle O \rangle)^2}
\end{equation}
only depends on the expectation value $\langle O \rangle = \bra{\psi_\varphi} O \ket{\psi_\varphi}$.

This expectation value can be evaluated by using the transformations $O (a^\dagger \pm b^\dagger) O^\dagger = \pm (a^\dagger \pm b^\dagger)$, and therefore $O c_k^\dagger(\varphi) O^\dagger =  c_k^\dagger(-\varphi)$ and $O d_k^\dagger(\varphi) O^\dagger = - d_k^\dagger(-\varphi)$. The expectation value
\begin{align}
\langle O \rangle
=&\multInt \frac{\prod_{i,j}^{m,m} \dd k_i \dd \tilde{k}_j \dd p_i \dd \tilde{p}_j}{(2\pi)^{4m} m!^4} 
A_{k_1 \cdots k_m}^* B_{\tilde{k}_1 \cdots \tilde{k}_m}^* A_{p_1 \cdots p_m} B_{\tilde{p}_1 \cdots \tilde{p}_m} (-1)^m
\notag \\
& \qquad \qquad		
\qquad 
\times  \bra{0}
c_{k_{1}}(\varphi)  \cdots c_{k_m}(\varphi) d_{\tilde{k}_{1}}(\varphi)  \cdots d_{\tilde{k}_m}(\varphi) 
c^\dagger_{p_1}(-\varphi) \cdots c^\dagger_{p_m}(-\varphi)  d^\dagger_{\tilde{p}_{1}}(-\varphi) \cdot  d^\dagger_{\tilde{p}_{m}}(-\varphi) 
\ket{0},
\notag \\	 \nonumber
\end{align}
can then be further evaluated by using the commutation relations
\begin{subequations}
	\begin{align}
	\left[c_p(\varphi), c_k^\dagger(-\varphi) \right] 
	=& \left[d_p(\varphi), d_k^\dagger(-\varphi) \right] = 2 \pi \delta(p-q) \cos \varphi, \\
	\left[c_p(\varphi), d_k^\dagger(-\varphi) \right] 
	=& \left[d_p(\varphi), c_k^\dagger(-\varphi) \right] = 2 \pi \delta(p-q) \ii \sin \varphi.
	\end{align}
\end{subequations}
Because the commutators between $c_k$ and $d_k$ do not vanish, all indices can become mixed and the expectation value yields
\begin{equation}
\langle O \rangle
=\sum_{l=0}^m (-1)^{l}  \sin^{2l}( \varphi) \cos^{2(m-l)}(\varphi) \binom{m}{l}^2 I_m^\klamm{l},
\end{equation}
where the integrals $I_m^\klamm{l}$ are the natural extension of \eqref{eq:Ims}, i.e., in $I_m^\klamm{l}$   $l$ indices are exchanged in the integral. We note, that $I_j^\klamm{l} = I_{m-j}^\klamm{l}$ such that one can reduce the number of calculations if they are necessary.

By observing that $\partial_\varphi \langle O \rangle \big\rvert_{\varphi = 0} = 0$ and $\langle O \rangle_{\varphi = 0} = I_m^\klamm{0} = 1$, the variance of the measured phase can be calculated from the second derivative,
\begin{equation}
\Delta \varphi^2 
= \lim_{\varphi \rightarrow 0} \frac{1- \langle O \rangle^2}{(\partial_\varphi \langle O \rangle)^2} 
=\lim_{\varphi \rightarrow 0} \frac{- 2 \langle O \rangle \partial_\varphi \langle O \rangle }{2 \partial_\varphi \langle O \rangle \partial_\varphi^2 \langle O \rangle} 
= \left( - \partial_\varphi^2 \langle O \rangle \big\rvert_0 \right)^{-1}.
\end{equation}
The second derivative around $\varphi \approx 0$ only contains the first two terms, for which the sine-terms vanish after the derivative, i.e. $- \partial_\varphi^2 \langle O \rangle \big\rvert_0 = 2 m (m I_m^\klamm{1} + I_m^\klamm{0})$. Therefore, we reach the QCRB locally around $\varphi \approx 0$,
\begin{equation}
\Delta \varphi^2  \big\rvert_{\varphi \approx 0} 
= \frac{1}{F_Q [\psi_\varphi^{AB}]}.
\end{equation}
We note that for single-mode states $I_N^\klamm{l} = 1$, which leads to the result derived in Reference~\cite{campos03a}. In that case the expectation value $\langle O \rangle = P_m \left[ \cos 2 \varphi \right]$ can be expressed in terms of Legendre Polynomials $P_m$. In that case, the second derivative $- \partial_\varphi^2 \langle O \rangle \big\rvert_0 = 4 P_m' [1]$ is calculated with the help of the well-known result $P_m' [1] = m(m+1)/2$.

\section{Transforming the bidirectional wavepacket into a unidirectional one} 
\label{app.transf}
Let us finally, give an example on how to merge the bidirectional wavepacket into a unidirectional one with the same metrological properties. The wavepacket emitted from each atomic ensemble reads
\begin{align}
\ket{\phi^{(N)}_A}
=\multIntbi \frac{ \dd k_1 ... \dd k_N}{(2\pi)^N N!}    A_{ \{ k\}}  a^\dagger_{k_1}  \dots {a}^\dagger_{k_{N}} \ket{0}.
\label{eq:multimodeII}
\end{align}
where the $k_i$ integrals run from $(-\infty,\infty)$. This means that the wavepacket is actually emitted in both left/right directions. It is possible however to join the left/right emission into a common wavepacket by joining both ends of the waveguide through a 50/50 beam splitter transformation. To make it more explicit, we can define $r_{k}/l_{k}$ for the $a_{k}$ modes propagating to the right/left ($k\gtrless 0)$, and rewrite the integral with integration ranges from $(0,\infty)$.
\begin{align}
\ket{\phi^{(N)}_A}
=\multIntone \frac{ \dd k_1 ... \dd k_N}{(2\pi)^N N!}    A_{ \{ k\}} \left[ r^\dagger_{k_1}  \dots {r}^\dagger_{k_{N-1}} {r}^\dagger_{k_{N}}+r^\dagger_{k_1}  \dots {r}^\dagger_{k_{N-1}}l^\dagger_{k_N}  + \dots + l^\dagger_{k_1}  \dots {l}^\dagger_{k_{N-1}} {l}^\dagger_{k_{N}}\right] \ket{0}.
\end{align}
The $A_{\{k\}}$ factorizes out from the sum because it has the symmetry $k_i\rightarrow -k_i$ since $\omega(k)\propto |k|$. Notice that now the sum can also be written as a product:
\begin{align}
\ket{\phi^{(N)}_A}
=\multIntone \frac{ \dd k_1 ... \dd k_N}{(2\pi)^N N!}    A_{ \{ k\}}  \left[\prod_{i=1}^N(l^\dagger_{k_i}+r^\dagger_{k_i}) \right]\ket{0}.
\label{eq:multimodeIII}
\end{align}
If the $l/r$ modes are used as inputs of a beam splitter such that the modes transform at the output ports C/D as $c^\dagger_{k_i}=\left(r^\dagger_{k_i}+l^\dagger_{k_i}\right)/\sqrt{2}$ and $d^\dagger_{k_i}=\left(-r^\dagger_{k_i}+l^\dagger_{k_i}\right)/\sqrt{2}$. Then:
\begin{align}
\ket{\phi^{(N)}_C}
=\multIntone \frac{ \dd k_1 ... \dd k_N}{(2\pi)^N N!}   2^{N/2} A_{ \{ k\}}  c^\dagger_{k_i}\ket{0}.
\label{eq:multimodeIV}
\end{align}
Since this state shares the same modal function, $A_{\{k\}}$, than the original one the metrological properties can be shown to be the same than the ones calculated in the main manuscript.

\section{Derivation of Recurrence Relation}\label{app:RecRel}

We now focus on the evaluation of the integral expression $I_{AB}^\klamm{1}$ in the case of the same multi-mode input states, that is, for $m = n = N/2$ and $A_{ \{k\}}=B_{ \{k\}}$. Since only the integral $I_{AB}^\klamm{1}$ is relevant for the discussion, from now on, and in the main manuscript we drop the superindex: $I_{AB}^\klamm{1}\equiv I_{AB}$. If the input state is a product state, that is, if $A_{ \{k\}}= \frac{1}{\sqrt{m!}} A_{k_1} A_{k_2} \cdot A_{k_m}$ factorizes, the $I_{2m} \equiv I_{AA}$ can be straightforwardly integrated in each $k_i$ and $\tilde{k}_i$ separately. This calculation yields $I_{2m} = 1$, so that the single mode result of $F_Q[\psi_\varphi^\mathrm{Fock}] = N (N+2)/2$ is recovered.

On the other hand, the coefficient of the photonic state emitted from a chain of quantum emitters along a waveguide does not factorize in this way, such that the evaluation of $I_{2m}$ requires additional effort. Because the multi-mode coefficients originate from the exponential decay of the emitters,
\begin{equation}
A_{ \{k\}} 
= (-\ii)^m \int_0^\infty \prod_{i} \dd t_i \
\eexp{\ii \sum_i k_i t_i}
\mathcal{T} \bra{0} O_{t_1} O_{t_2} \cdots O_{t_m} \ket{\psi_m},
\end{equation}
where $O_t \equiv O(t) = \sqrt{\Gamma_\oned}  \eexp{\ii H_\mathrm{eff} t} S_{ge} \eexp{-\ii H_\mathrm{eff} t}$ with the effective Hamiltonian $H_\mathrm{eff} = (\Delta - \ii \frac{\Gamma^*}{2}) S_{ee} - \ii \frac{\Gamma_\oned}{2} S_{eg}S_{ge}$ acts on the symmetric Dicke states $\ket{\psi_m} = \frac{1}{m!} \binom{N}{m}^{-1/2} S_{eg}^m \ket{0}$. The action of the time ordering operator $\mathcal{T}$ on commuting operators is defined as $\mathcal{T} O_{t_1} O_{t_2} = \theta(t_1 - t_2) O_{t_1} O_{t_2} + \theta(t_2 - t_1) O_{t_2} O_{t_1}$.
Using this expression for the coefficients $A_{\{k\}}$ the integrals in momentum space can be transformed to integrals in time,
\begin{align}
I_{2m}
= \frac{1}{m!^2}
\int_0^\infty \prod_{i,j} \dd t_i \dd s_j \mathcal{T} &
\bra{0} O_{t_1} O_{t_2} \cdots O_{t_m} \ket{\psi_m}^*
\bra{0} O_{s_1} O_{s_2} \cdots O_{t_m} \ket{\psi_m}^*
\nonumber \\
& \times \bra{0} O_{s_1} O_{t_2} \cdots O_{t_m} \ket{\psi_m}
\bra{0} O_{t_1} O_{s_2} \cdots O_{t_m} \ket{\psi_m}.
\end{align}
Notice, that in the correlation functions one index is exchanged, in analogy with the expressions in momentum space, and that the integral is symmetric with respect to the remaining $t_i$/$s_j$ indices.

The integral can be evaluated recursively by picking a time ordering and integrating over the latest time $\tau \geq \max_{\neq \tau} \{t_i, s_i\} \equiv T$, and repeating this step on the next integral. The exponential decay then gives rise to the simple form of $\int_{T}^\infty \eexp{- c \tau} = \frac{1}{c}  \eexp{- c T}$ if $\Re(c)>$0. Using these results, one can define three structurally different integrals, depending on whether one has already integrated over one or both of the special (i.e., exchanged) indices $t_1$ or $s_1$,
\begin{subequations}
	\begin{align}
	F_{ij}^\klamm{2} =& 
	\int \prod_{i',j'} \dd t_{i'} \dd s_{j'} \mathcal{T} 
	\eexp{- c_{ij}^\klamm{2}\max \{t_{i'}, s_{j'} \}}
	\bra{\psi_{m-1-i}} O_{t_1} O_{t_2} \cdots O_{t_{i+1}} \ket{\psi_m}^*
	\bra{\psi_{m-1-j}} O_{s_1} O_{s_2} \cdots O_{s_{j+1}} \ket{\psi_m}^*
	\nonumber \\
	& \qquad \qquad \qquad \qquad \qquad
	\times \bra{\psi_{m-1-i}} O_{s_1} O_{t_2} \cdots O_{t_{i+1}} \ket{\psi_m}
	\bra{\psi_{m-1-j}} O_{t_1} O_{s_2} \cdots O_{s_{j+1}} \ket{\psi_m},
	\\
	F_{ij}^\klamm{1} =& 
	\int \prod_{i',j'} \dd t_{i'} \dd s_{j'} \mathcal{T} 
	\eexp{- c_{ij}^\klamm{1}\max \{t_{i'}, s_{j'} \}}
	\bra{\psi_{m-1-i}} O_{t_1} O_{t_2} \cdots O_{t_{i+1}} \ket{\psi_m}^*
	\bra{\psi_{m-j}} O_{s_2} \cdots O_{s_{j+1}} \ket{\psi_m}^*
	\nonumber \\
	& \qquad \qquad \qquad \qquad \qquad
	\times \bra{\psi_{m-i}} O_{t_2} \cdots O_{t_{i+1}} \ket{\psi_m}
	\bra{\psi_{m-1-j}} O_{t_1} O_{s_2} \cdots O_{s_{j+1}} \ket{\psi_m},
	\\
	F_{ij}^\klamm{0} =& 
	\int \prod_{i',j'} \dd t_{i'} \dd s_{j'} \mathcal{T} 
	\eexp{- c_{ij}^\klamm{0}\max \{t_{i'}, s_{j'} \}}
	\bra{\psi_{m-i}} O_{t_2} \cdots O_{t_{i+1}} \ket{\psi_m}^*
	\bra{\psi_{m-j}} O_{s_2} \cdots O_{s_{j+1}} \ket{\psi_m}^*
	\nonumber \\
	& \qquad \qquad \qquad \qquad \qquad
	\times \bra{\psi_{m-i}} O_{t_2} \cdots O_{t_{i+1}} \ket{\psi_m}
	\bra{\psi_{m-j}} O_{s_2} \cdots O_{s_{j+1}} \ket{\psi_m},
	\end{align}
\end{subequations}

The integrals only run over the remaining time variables $\{ t_{i'} \}$ and $\{ s_{j'} \}$ and we have introduced the exponents $c_{ij}^\klamm{2} = \gamma_{m-1-i} + \gamma_{m-1-j}$, $c_{ij}^\klamm{0} = \gamma_{m-i} + \gamma_{m-j}$, and $c_{ij}^\klamm{1} =  (c_{ij}^\klamm{2} + c_{ij}^\klamm{0})/ 2$. The decay rates are given by $\gamma_j = j (N-j+1) \Gamma_\oned$ defined through $\Gamma_\oned S_{eg} S_{ge} \ket{\psi_j} = \gamma_j \ket{\psi_j}$. Note that these integrals always converge because $c_{ij}^\klamm{2/1/0}>0$.

\begin{figure}
	\centering
	\includegraphics[width=0.8\textwidth]{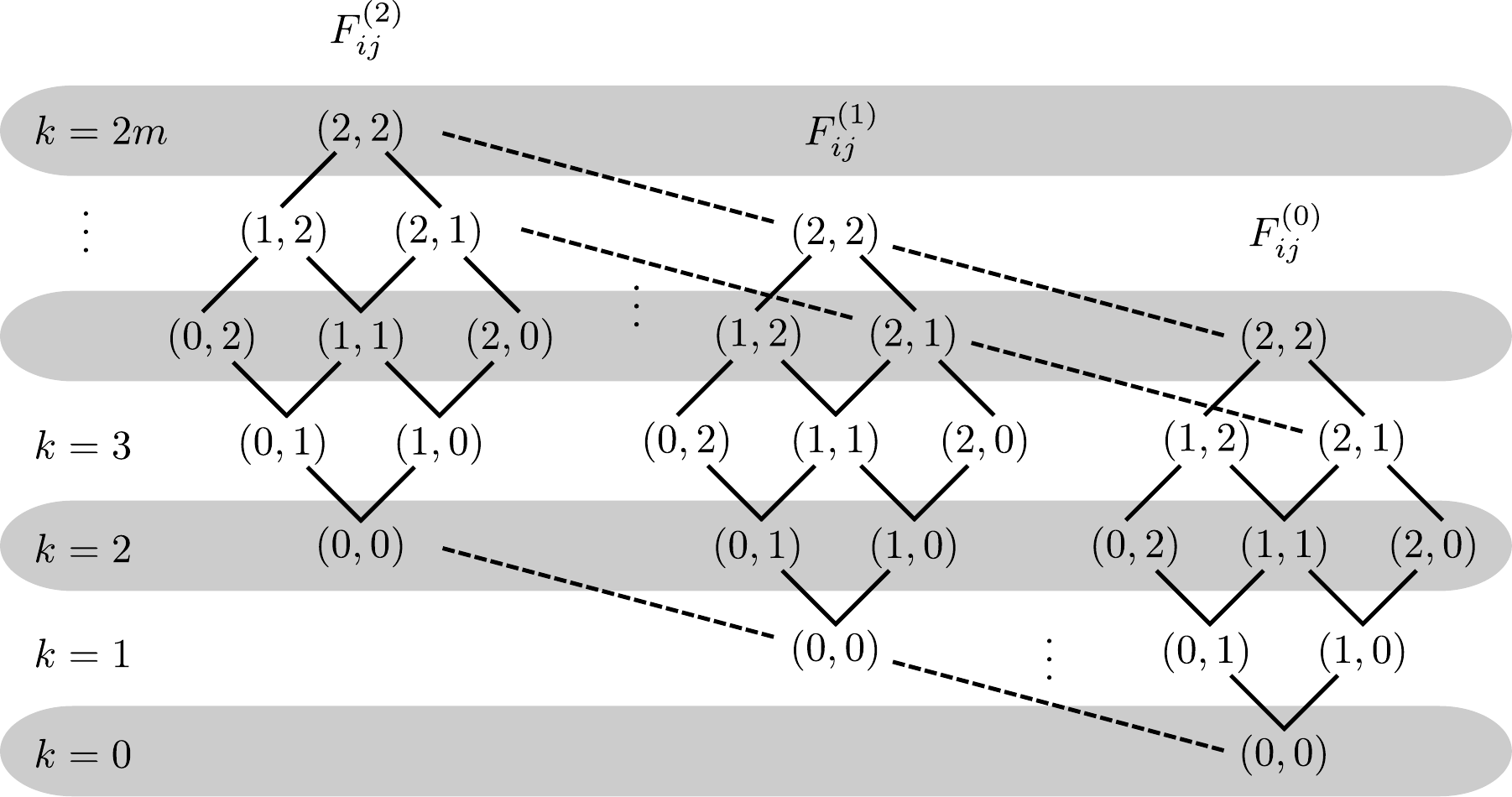}
	\caption{The recurrence relation of $F_{ij}^\klamm{2/1/0}$ to calculate $I_m$ can be represented pictorially, here on the example of $m=3$. The solid lines represent the terms of the recurrence relation in between every group $F^\klamm{n} \rightarrow F^\klamm{n}$, whereas the dashed lines correspond to the terms $F^\klamm{2} \rightarrow F^\klamm{1}$ and  $F^\klamm{1} \rightarrow F^\klamm{0}$. By grouping the elements in terms of the number of excitations, or equivalently the number of remaining time integrals, one can evaluate the recurrence relation efficiently.}
	\label{fig1:Recursion}
\end{figure}

By integrating over the latest time, one can remove one operator $O_{t_i}$ or $O_{s_j}$ from the above expressions until one ends up with $F_{00}^\klamm{0} = 1$. This motivates the fact that the integral 
\begin{equation}
I_{2m} 
= \frac{1}{m!^2} F_{m-1,m-1}^\klamm{2}
\end{equation}
can be evaluated by a recurrence relation (see also Figure \ref{fig1:Recursion}). Let us understand the structure of the recurrence relation on the example of $F_{ij}^\klamm{2}$. If the largest time is one with a regular index $t_2$, ... $t_{i+1}$ (for which there are $i$ possibilities), we use the fact that
\begin{equation}
\bra{\psi_{m-1-i}} O_{t_{i+1}} 
= \sqrt{\gamma_{m-i}} \eexp{-(\gamma_{m-i} - \gamma_{m-i-1}) t_{i + 1}/2} \bra{\psi_{m-1-i}}.
\end{equation}
This term appears twice such that the integral gives a prefactor $\frac{\gamma_{m-i}}{c_{ij}^\klamm{2} + (\gamma_{m-i} - \gamma_{m-i-1})} = \frac{\gamma_{m-i}}{c_{i-1,j}^\klamm{2}}$. The remaining integral is then of the form $F_{i-1,j}^\klamm{2}$. The same holds if the largest time is one of $s_2$, ... $s_{j+1}$. If the largest time is $s_1$ (or equivalently $t_1$), then after the integration over this variable, the remaining integral is of the form $F_{ij}^\klamm{1}$. By carefully calculating all these steps, we find the recurrence relation
\begin{subequations}\label{eq:Recurrence}
	\begin{align}
	F_{ij}^\klamm{2} 
	=& 
	i \frac{\gamma_{m-i}}{c_{i-1,j}^\klamm{2}} F_{i-1,j}^\klamm{2}
	+ j \frac{\gamma_{m-j}}{c_{i,j-1}^\klamm{2}} F_{i,j-1}^\klamm{2}
	+ 2 \frac{\sqrt{\gamma_{m-i} \gamma_{m-j}}}{c_{i,j}^\klamm{1}} F_{i,j}^\klamm{1},
	\\
	F_{ij}^\klamm{1} 
	=& i \frac{\sqrt{\gamma_{m-i} \gamma_{m-i+1}}}{c_{i-1,j}^\klamm{1}} F_{i-1,j}^\klamm{1}
	+ j \frac{\sqrt{\gamma_{m-j} \gamma_{m-j+1}}}{c_{i,j-1}^\klamm{1}} F_{i,j-1}^\klamm{1}
	+ \frac{\sqrt{\gamma_{m-i} \gamma_{m-j}}}{c_{i,j}^\klamm{0}} F_{i,j}^\klamm{0},
	\\
	F_{ij}^\klamm{0} 
	=& 
	i \frac{\gamma_{m-i+1}}{c_{i-1,j}^\klamm{0}} F_{i-1,j}^\klamm{0}
	+ j \frac{\gamma_{m-j+1}}{c_{i,j-1}^\klamm{0}} F_{i,j-1}^\klamm{0},
	\\
	F_{00}^\klamm{0} 
	=& 1.
	\end{align}
\end{subequations}

The trick to evaluating this recurrence relation efficiently is to group elements of the same excitation subspace $0 \leq k \leq 2 m$ as in Figure \ref{fig1:Recursion}. Elements of this subspace are, for example, $F_{ij}^\klamm{2}$ satisfying $i + j + 2 = k$ and $0\leq i,j \leq m-1 $. By applying one recursive step starting from $k=0$, in which only $F_{00}^\klamm{0}=1$ lies, one moves to a subspace with one excitation more $k \rightarrow k+1$ until $k = 2m$ is reached. This subspace only contains the desired term $F_{m-1,m-1}^\klamm{2}$. For better numerical results it is also recommendable to remove the factors of $i$ and $j$ by substituting $F_{ij}^\klamm{n} = i! j! \tilde{F}_{ij}^\klamm{n}$.

\section{Robustness to Errors}\label{App.robust}

In this Section, we estimate how the different error sources affect our protocol, deriving the conditions under which they can be neglected. In particular, we study the impact of i) free-space spontaneous emission, ii) finite propagation length of the modes, iii) retardation effects due to finite group velocity, iv) different coupling to the waveguide of the two emitter ensembles, v) time delay between the different wavepackets, and vi) photon loss in the interferometer.

\subsection{Impact of Emission into Free Space}
\label{App.ErrorsI}
One of the greatest sources of decoherence in state-of-art waveguide QED systems is the possibility of emitting to other modes different from the relevant waveguide one. We embed all these processes into a single decay rate, $\Gamma^*$, and describe through an individual Lindblad decay terms as follows:
\begin{equation}
\label{eqSM:Linblad}
\mathcal{L}_*[\rho]=\frac{\Gamma^*}{2}\sum_{n=1}^N\left(2\sigma_{ge}^n\rho\sigma_{eg}^n-\sigma_{ee}^n\rho-\rho\sigma_{ee}^n\right)
\end{equation}

With this extra term, the effective non-hermitian Hamiltonian governing the atomic state evolution contains now two contributions: the collective and individual decay terms, which read:
\begin{align}
\label{eq:Heff}
H_\mathrm{eff}=-i\left(\frac{\Gamma_\oned}{2}S_{eg}S_{ge}+\frac{\Gamma^*}{2}\sum_{n} \sigma_{ee}^n\right)\,,
\end{align}
as well as the quantum jumps evolution:
\begin{align}
J[\rho]&=J_\oned[\rho]+J_*[\rho]\,,\\
J_\oned[\rho]&=\Gamma_\oned S_{ge}\rho S_{eg}\,,\\
J_{*}[\rho]&=\Gamma^* \sum_{n}\sigma^n_{ge}\rho \sigma_{eg}^n\,.
\end{align}

The formal evolution of $\rho(t)$ can be formally integrated as a sum of different contributions: $\rho(t)=\sum_{j}\rho_j(t)$ depending on the number of quantum jumps, denoted by $j$, that has occurred during the evolution. In particular, the different $\rho_j(t)$ can be formally computed as:
\begin{align}
\label{eq:formal}
\rho_0(t)&=S(t,t_0)\rho(t_0)\,,\\
\rho_{j\ge 1}(t)&=\int_0^t dt_1 S(t,t_1)J[\rho_{j-1}(t_1)]\,, \label{eq:formal111}
\end{align}
where we have defined the following operator:  $S(t_2,t_1)[\rho]=e^{-i H_\mathrm{eff} t}\rho e^{i H_\mathrm{eff}^\dagger t}$ which gives the evolution under the non-hermitian Hamiltonian. Since we assume an initial state $\ket{\Psi(0)}=\ket{e}^{\otimes N}$ and we are only interested in the probability of decaying to $\ket{g}^{\otimes N}$ only trough collective quantum jumps (denoted as $p$ in the main text), we restrict our attention to the dynamics of the collective atomic states with $m$ excitations, that is, $\ket{m}\propto S_{ge}^{N-m}\ket{\Psi(0)}$, that we denote as:
\begin{equation}
\label{eq:pm}
P_m(t)=\bra{m}\rho(t)\ket{m}\,.
\end{equation}

Using this notation $P_0(t\rightarrow\infty)\equiv p$. Since only collective quantum jumps participate in the evolution of $P_m(t)$, their dynamics can be calculated straightforwardly from Eqs.~\ref{eq:formal}-\ref{eq:formal111}. First, note that the non-Hermitian Hamiltonian only connect states with the same number of excitations, such that:
\begin{align}
\label{eq:ev}
&S(t_2,t_1)\left[\ket{m}\bra{m}\right]=\ket{m}\bra{m}e^{-\left[\Gamma_\oned m(N-m+1)+m\Gamma^* \right](t_2-t_1)}
\end{align}

For example, the evolution of the higher excited state is simply given by:
\begin{equation}
\label{eq:p0}
P_N(t)=e^{-\left(\Gamma_\oned+\Gamma^*\right)N t}\,.
\end{equation}

From here, the evolution of the $p_{m<N}(t)$ can be calculated recursively using Eq.~\ref{eq:formal111}:
\begin{equation}
P_m(t)=m(N-m++1)\Gamma_\oned \int_0^t d t_1 e^{-\left[(m-1)(N-m+2)\Gamma_\oned+(m-1)\Gamma^*\right](t-t_1)} P_{m+1}(t)\,.
\end{equation}

Using these formulas one can calculate the dynamics of $P_m(t)$ for all $m$ and set of parameters, $N$, $\Gamma_\oned$ and $\Gamma^*$. To gain intuition from the decay process, we start calculating $P_m(t)$ for a situation with $\Gamma^*=0$, that we show in different colors in Fig.~\ref{figSM:loss}(a) for a situation with $N=20$ QEs. We start observing a collective decay from the highly excited state (in red) $m=N=20$, as the lower excited levels starts building up population until it gets accumulated in $m=0$. From this figure, it may look as if the transient time through the higher excited states was faster than in the smaller ones. However, by looking into the averaged time population:
\begin{equation}
\label{eqSM:pm}
\bar{P}_m=\int_0^\infty dt P_m(t)\,.
\end{equation}
which we plot in the inset of the Figure, we observe that in fact the average time spent in each of the levels distribute symmetrically around $m=N/2+1$. Thus, when considering $\Gamma^*\neq 0$, the main source of errors will come from the upper part of the ladder $m\approx N$, since the decay rate into free-space is proportional to the number of excitations $\sim m \Gamma^*$.

\begin{figure*}[tb]
	\centering
	\includegraphics[width=0.8\linewidth]{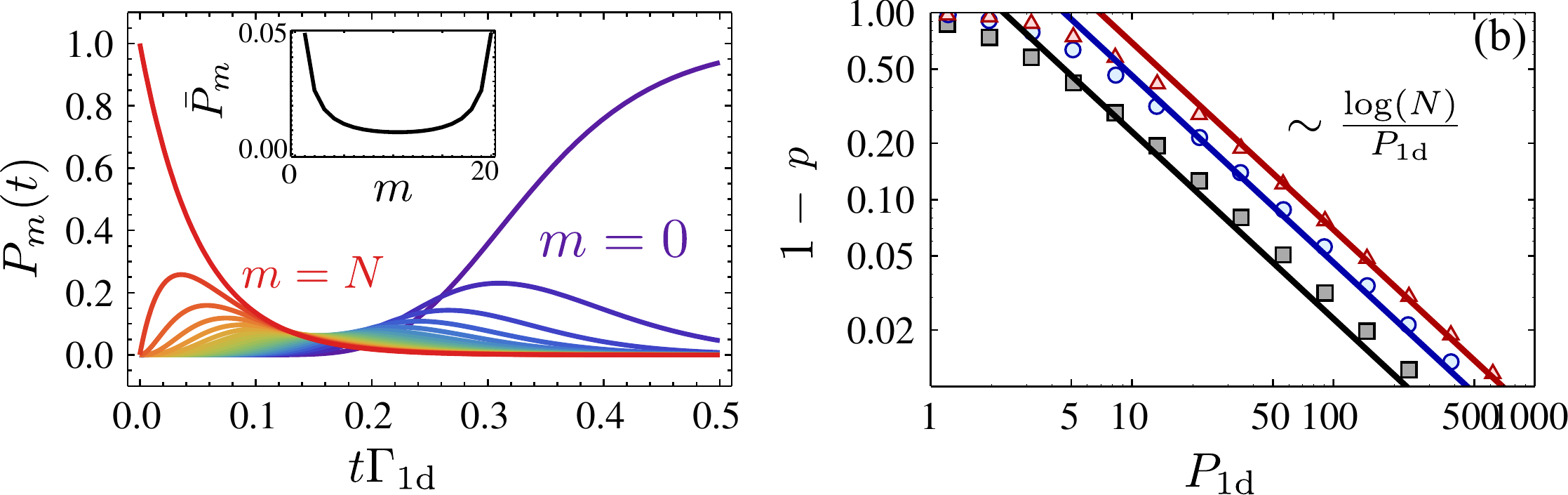}
	\caption{(a) $P_m(t)$ for a situation with $N=20$ and $P_\oned$ for different $m$ ranging from $m=N=20$ (red) to $m=0$ purple. Inset: Integrated population $\bar{P}_m=\int_0^\infty P_m(t)dt$ with the parameters. (b) $1-P_0(t\rightarrow\infty)$ [$1-p$] as a function of $P_\oned$ for $N=10$ (black squares), $N=100$ (blue spheres) and $N=1000$ (red triangles).}
	\label{figSM:loss}
\end{figure*}

In Fig.~\ref{figSM:loss}(b) we show the effect of $\Gamma^*\neq 0$ on $p$, which is the relevant parameter to estimate the lower bound of the QFI given in the main text. In particular, we plot the scaling of $1-p$ as a function of $P_\oned=\frac{\Gamma_\oned}{\Gamma^*}$ for several $N$'s as depicted in the legend. We observe that the exact calculation of $1-p$ obtains the same scaling, $\log(N)/P_\oned$, as we show in the main text with a simplified description of the losses. 

This error scaling can also be obtained by estimating the timescale of the superradiant decay as the sum of the different decay timescales of $P_m(t)$, which leads to:
\begin{equation}
\label{eq:p}
\tau_\mathrm{SR}\approx \sum_{j=1}^N \frac{1}{\Gamma_\oned j(N-j+1)}\sim \frac{\log(N)}{N\Gamma_\oned}
\end{equation}

With this timescale, one can easily upper bound the error of $1-p$ by multiplying this timescale by the maximum error rate, $N\Gamma^*$, from the higher excited state. This results into an upper bound
\begin{equation}
1-p\le N\Gamma^*\tau_\mathrm{SR}\approx \frac{\log(N)}{P_\oned}\,,
\end{equation}
which has the same scaling as the one observed in the numerical simulations.

\subsection{Finite lifetime of waveguide modes}

Another possible source of decoherence is the finite lifetime of the photonic waveguide modes, which appears due to absorption or imperfections in the material which leads to scattering into other modes. These photonic losses affect the metrological properties during and after the $N$-photon emission. The effect of the losses after the wavepacket has been emitted can be considered as noise within the interferometric process, which have been well studied in the literature~\cite{demkowicz-dobrzanski15}, leading to a loss of Heisenberg scaling for large $N$. Since this is a common limitation of all metrological protocols, we focus on the effect of photon losses during the emission of the $N$-photon wavepacket. 

During the emission of the waveguide, the finite lifetime of waveguide modes induce a finite propagation of waveguide modes, $L_\mathrm{prop}$, which spoils the collective behaviour of the emitter interactions as follows:
\begin{equation}
\Gamma_{m,n}=\Gamma_\oned e^{-|x_n-x_m|/L_\mathrm{prop}}\,.
\end{equation}

To be able to neglect this correction, the propagation length of the modes has to be much larger than the system size, that is, $L_\mathrm{prop}\gg N\lambda_a$, where we have assumed a separation between atoms of the order of $\lambda_a$, required to have the perfect collective behaviour. The propagation length of the modes is approximately given by~\cite{gonzaleztudela15a}:
\begin{equation}
\frac{L_\mathrm{prop}}{\lambda_a}\approx \frac{Q}{2 n_g}\,,
\end{equation}
where $Q$ is the experimental quality factor of the waveguide modes, whereas $n_g$ is the so-called group index, which measures the reduction of the speed of light within the waveguide. Thus, the inequality that must be satisfied is that:
\begin{equation}
\frac{Q}{2 n_g}\gg N\,.
\end{equation}

State-of-the-art numbers with SiN waveguides~\cite{goban13a} are $Q\approx  10^6$ and $n_g\approx 10$, which gives $L_\mathrm{prop}/\lambda_a\approx 5\times 10^4$. Since this size is even larger than typical waveguide lengths, this correction will be typically small.

\subsection{Retardation effects: validity of the Markov approximation}

All the calculations shown in this manuscript, including the one of the spectral shape of the wavepacket, $A_{\{q\}}$, are performed by using a Born-Markov master equation describing the atomic dynamics as written in the main text. The underlying assumption of this equation is that the bath timescales are faster than the emitter ones. In particular, the emergence of superradiant behaviour as predicted by Dicke superradiance requires that the propagation time of the photons between all the emitters is faster than the fastest emitter timescale. The maximum propagation time for a system with $N$ emitters is given by:
\begin{equation}
\label{eq:tprop}
\tau_\mathrm{prop}=\frac{N\lambda_a}{v_g}\,.
\end{equation}
where $v_g=c/n_g$ is the group velocity of the photons in the waveguide. The fastest atomic timescale occurs in the middle of the Dicke Ladder, where the decay rate scales with $\sim \Gamma_\oned N^2/4$. Thus, the condition that must be satisfied is that:
\begin{equation}
\tau_\mathrm{prop}\ll \frac{4}{\Gamma_\oned N^2}\rightarrow N^3\ll \frac{4 c}{n_g \lambda_a \Gamma_\oned}
\end{equation}

Using state-of-art numbers of $\Gamma_\oned\sim 2\pi\times 6$ MHz, $n_g\approx 10$ and $\lambda_a=300$ nm, we find $4 c/(n_g \lambda_a \Gamma_\oned)\sim 10^7$, which implies $N<200$. Moreover, by making use of a Raman transition one can decrease $\Gamma_\oned$, while at the same time attenuating $\Gamma^*$ such that $P_\oned$ remains fixed.

\subsection{Different Purcell Factors between wavepackets}

Until now we have assumed that the ensembles generating the multi-mode state $\ket{\phi_A^{N}}$ are coupled with the same decay rate, $\Gamma_\oned$, to the waveguide modes. Let us now assume they are different, that is, that they are coupled with $\Gamma_\oned$ and $\Gamma'_\oned$, respectively. Note, that this does not affect the norm of the state, but it does change the integral $I_N$. This can still be evaluated by a similar recurrence relation as in \eqref{eq:Recurrence}. The only difference is that in every numerator one has to replace $\gamma_j \rightarrow \sqrt{\gamma_j \gamma'_j}$ and in every denominator, that is in every $c_{ij}^\klamm{l}$, $\gamma_j \rightarrow \frac{1}{2} \left(\gamma_j + \gamma'_j\right)$. Because $\gamma'_j = \gamma_j \frac{\Gamma'_\oned}{\Gamma_\oned}$ every step of the recurrence relation gets an additional factor of $\frac{\sqrt{\Gamma'_\oned / \Gamma_\oned}}{\frac{1}{2}\left(1 +  \Gamma'_\oned / \Gamma_\oned \right)}$. As there are $N$ steps in the recurrence relation,
the integral $I_N$ has to be replaced by
\begin{equation}
I_{N,\Delta\Gamma_\oned}=\left( \frac{2 \sqrt{\Gamma_\oned \Gamma'_\oned}}{\Gamma_\oned + \Gamma'_\oned} \right)^N I_N
= I_N \left(1 - \frac{N}{8} \left(\frac{\Delta\Gamma_\oned}{\Gamma_\oned}\right)^2 + \mathcal{O}\left[\left(\frac{\Delta\Gamma_\oned}{\Gamma_\oned}\right)^3\right] \right),
\end{equation}
where $\Delta\Gamma_\oned = (\Gamma_\oned-\Gamma_\oned')$.

\subsection{Time Delay between Wavepackets}

Another deviation from the ideal situation appears if the wavepackets emitted from the first/second ensemble does not arrive simultaneously to the beam splitter. This can occur if either the collective $\pi$-pulse exciting the ensembles is not perfectly simultaneous or the travelling path between the two wavepackets is not exactly matched. In both cases, they will give rise to a time delay, $\tau$, between the two wavepackets. This time delay enters in the integral $I_N$ as follows:
\begin{equation}
I_{N,\tau}
= \multInt \frac{\prod_{i=1}^{n}\dd k_i \dd \tilde{k}_i}{(2\pi)^{2n} n! n!} 
A^*_{k_1,...,k_n} A^*_{\tilde{k}_1,...,\tilde{k}_n} 
A_{\tilde{k}_1,k_2,...,k_n}A_{k_1,\tilde{k}_2...,\tilde{k}_n}
\eexp{-\ii \tau (k_1 - \tilde{k}_1)},
\end{equation}
where $n = N/2$ and $\tau$ the delay between the wavefronts. By transforming this integral in momentum space to an integral in time space, we find that it is equivalent to
\begin{align}
I_{N,\tau}
= \frac{1}{n!^2}
\int_0^\infty \prod_{i} \dd t_i \int_{-\tau}^{\infty} \prod_j \dd s_j &
\theta(t_1-\tau) \eexp{-\gamma_n \tau}
\bra{0}  \mathcal{T} O_{t_1} O_{t_2} \cdots O_{t_m} \ket{\psi_n}^*
\bra{0}  \mathcal{T} O_{s_1} O_{s_2} \cdots O_{t_m} \ket{\psi_n}^*
\nonumber \\
& \times \bra{0}  \mathcal{T} O_{s_1} O_{t_2} \cdots O_{t_m} \ket{\psi_n}
\bra{0}  \mathcal{T} O_{t_1} O_{s_2} \cdots O_{t_m} \ket{\psi_n},
\end{align}
where $\theta(x)$ is the Heaviside function.
One can find a similar recurrence relation, which one can lower bound by noting that $\int_{T}^\infty \dd t\ \theta(t - \tau) \eexp{-c t} \geq \theta(T-\tau) \int_{T}^\infty \dd t\ \eexp{-c t}$. This means, that the Heaviside function appears in every remaining integral after the integral over $t_1$ has been performed. The final integral is then either of the form $\int_{0}^\infty \dd t\ \theta(t-\tau) \eexp{-\gamma_n t}$ or $\int_{-\tau}^\infty \dd s\ \theta(s-\tau) \eexp{-\gamma_n s}$, which both yield an additional factor of $\eexp{-\gamma_n \tau}$ in addition to the integral one would have to perform without the time delay. Therefore, the integral $I_{N,\tau}$ is lower bounded by 
\begin{equation}
I_{N,\tau}
\geq \eexp{-2 \gamma_n \tau} I_N= \eexp{-2 \gamma_{N/2} \tau} I_N
\approx I_N \left( 1 - N\Gamma_\oned \tau + \mathcal{O}\left[(N\Gamma_\oned \tau )^2 \right]\right).
\end{equation}

One can compare this to the single-mode result, for which one obtains
\begin{equation}
I_{N,\tau}
= \eexp{-\gamma_1 \tau}
\approx 1 - N \Gamma_\oned \tau/2 +\mathcal{O}\left[(N\Gamma_\oned \tau )^2 \right]\,.
\end{equation}

\subsection{Photon loss in the interferometer}

In this section, we characterise the first order corrections due to photon loss in one arm of the interferometer. This is described by a beam splitter that mixes the modes $b_k$ with an external mode ($e_k$) in the vacuum state with a reflection coeffficient $\sqrt{\eta}$. That is, 
\begin{align}
b_k \longrightarrow \sqrt{1-\eta} b_k + \sqrt{\eta}e_k^{\dagger}. 
\label{mixingb}
\end{align}
We also focus in the case $m=n=N/2$ and for twin-states $A=B$.
Due to the mixing \eqref{mixingb} the state $\ket{\psi_\varphi}$ changes as 
\begin{align}
\ket{\psi_\varphi} \longrightarrow \ket{\psi_\varphi}^{\rm noise}
= &\multInt \frac{\prod_{i,j=1}^{m,n} \dd k_i \dd \tilde{k}_j }{(2\pi)^{2n}  n!^2}
A_{k_1 \ldots k_n} A_{\tilde{k}_1 \ldots \tilde{k}_n} \nonumber\\
& \prod_j  \frac{1}{2}\left(e^{-\ii \varphi/2}a_{k_j}^\dagger + \eexp{\ii \varphi/2} (\sqrt{1-\eta}b_{k_j}^\dagger+\sqrt{\eta}e_{k_j}^\dagger) \right)\left(-e^{-\ii \varphi/2}a_{\tilde{k}_j}^\dagger + \eexp{\ii \varphi/2} (\sqrt{1-\eta}b_{k_j}^\dagger+\sqrt{\eta}e_{k_j}^\dagger)  \right)  \ket{0}.
\end{align}
After tracing out over the undesired mode $e_k$ $\forall k$ the state can be written as,
\begin{align}
\rho= \Tr_e \left ( \ket{\psi_\varphi}^{\rm noise} \bra{\psi_\varphi}^{\rm noise}\right) = \sum_{j=0}^{N} p_j \sigma^{(j)}
\end{align}
where $\sigma^{(j)}$ is a state that has lost $j$ photons into the modes $e_k$. Because each state has a different photon number, it follows
\begin{align}
F_Q [ \rho ] = \sum_{j=0}^{N} p_j F_Q [ \sigma^{(j)}].\label{eq:qfloss}
\end{align}
The computation of $F_Q [ \sigma^{(j)}]$ is challenging because the $\sigma^{(j)}$'s with $1\leq j<N$ are mixed states when the state is multimode. 
The techniques developed here only allow us for dealing with pure states, and we leave the development of techniques to compute the QFI of mixed multimode states as an interesting problem for the future. 
Here, instead,  we focus in the regime of small losses and characterise the first order corrections to the QFI due to photon loss.

We now focus on the limit $\eta\ll 1$, and in what follows we will only keep first order corrections in $\eta$ (so that $\approx$ stands for equality up to corrections of order $\mathcal{O}(\eta^2)$).  We focus on the state $\sigma^{(0)}$, which is a pure state $\sigma^{(0)}=\ket{\psi_{\varphi}^{(0)}} \bra{\psi_{\varphi}^{(0)}}$ as  no photons have been lost. We have, $p_0 \sigma^{(0)}=\ket{\tilde{\psi}_{\varphi}^{(0)}} \bra{\tilde{\psi}_{\varphi}^{(0)}}$ with the non-normalised state
\begin{align}
\ket{\tilde{\psi}_{\varphi}^{(0)}}= \multInt \frac{\prod_{i,j=1}^{n} \dd k_i \dd \tilde{k}_j }{(2\pi)^{2n} n!^2}
A_{k_1 \ldots k_n} A_{\tilde{k}_1 \ldots \tilde{k}_n} 
\tilde{c}^\dagger_{k_1}  \dots \tilde{c}^\dagger_{k_n}   \tilde{d}^\dagger_{\tilde{k}_1} \dots \tilde{d}^\dagger_{\tilde{k}_n} \ket{0}.
\end{align}
where we defined 
\begin{align}
&\tilde{c}^\dagger_{k}=\frac{1}{\sqrt{2}}\left(e^{-\ii \varphi/2}a_{k_j}^\dagger + \eexp{\ii \varphi/2} \sqrt{1-\eta}b_{k_j}^\dagger \right) \approx c^{\dagger}_{k_j}-\frac{\eta \eexp{\ii \varphi/2}}{2\sqrt{2}}b_{k_j}^{\dagger}
\nonumber\\
&  \tilde{d}^\dagger_{k}=\frac{1}{\sqrt{2}}\left(-e^{-\ii \varphi/2}a_{k_j}^\dagger + \eexp{\ii \varphi/2} \sqrt{1-\eta}b_{k_j}^\dagger \right) \approx d^{\dagger}_{k_j}-\frac{\eta \eexp{\ii \varphi/2}}{2\sqrt{2}}b_{k_j}^{\dagger}.
\end{align}
Expanding $\ket{\tilde{\psi}_{\varphi}^{(0)}}$ at first order in $\eta$ we obtain,
\begin{align}
\ket{\tilde{\psi}_{\varphi}^{(0)}} &\approx  \multInt \frac{\prod_{i,j=1}^{n} \dd k_i \dd \tilde{k}_j }{(2\pi)^{2n} n!^2}
A_{k_1 \ldots k_n} A_{\tilde{k}_1 \ldots \tilde{k}_n} 
\bigg( c^\dagger_{k_1}  d^\dagger_{\tilde{k}_1}
- \frac{n \eta\eexp{\ii \varphi/2}}{2\sqrt{2}}(b_{k_1}^{\dagger}d_{\tilde{k}_1}^{\dagger}+c_{k_1}^{\dagger}b_{\tilde{k}_1}^{\dagger}) \bigg) c^\dagger_{k_2}  \dots c^\dagger_{k_n}   d^\dagger_{\tilde{k}_2} \dots d^\dagger_{\tilde{k}_n}  \ket{0}
\nonumber\\
&=  \multInt \frac{\prod_{i,j=1}^{n} \dd k_i \dd \tilde{k}_j }{(2\pi)^{2n} n!^2}
A_{k_1 \ldots k_n} A_{\tilde{k}_1 \ldots \tilde{k}_n} 
\bigg( c^\dagger_{k_1}  d^\dagger_{\tilde{k}_1}
- \frac{n \eta}{4}(2c_{k_1}^{\dagger}d_{\tilde{k}_1}^{\dagger}+d_{k_1}^{\dagger}d_{\tilde{k}_1}^{\dagger}	 +c_{\tilde{k}_1}^{\dagger}c_{k_1}^{\dagger} \bigg) c^\dagger_{k_2}  \dots c^\dagger_{k_n}   d^\dagger_{\tilde{k}_2} \dots d^\dagger_{\tilde{k}_n}  \ket{0}
\end{align}
where we used the symmetry of $ A_{k_1 \ldots k_n}$ and $B_{\tilde{k}_1 \ldots \tilde{k}_n} $ over permutations. By a similar calculation of the ones performed in the previous sections, and recalling that $c_k$ ($d_k$) commutes with  $d_k^{\dagger}$ ($d_k$), one obtains,
\begin{align}
\braket{\tilde{\psi}_{\varphi}^{(0)}}{\tilde{\psi}_{\varphi}^{(0)}} \approx  1 -n\eta.
\end{align}
Hence, we have that,
\begin{align}
p_0 \approx 1-n\eta
\end{align}
and $\sigma^{(0)}=\ket{\psi_{\varphi}^{(0)}} \bra{\psi_{\varphi}^{(0)}}$ with 
\begin{align}
\ket{\psi_{\varphi}^{(0)}}& \approx \frac{1}{\sqrt{1-n\eta}}\ket{\tilde{\psi}_{\varphi}^{(0)}}\approx \left(1+\frac{n\eta}{2}\right)\ket{\tilde{\psi}_{\varphi}^{(0)}} 
\nonumber\\
&= \multInt \frac{\prod_{i,j=1}^{n} \dd k_i \dd \tilde{k}_j }{(2\pi)^{2n} n!^2}
A_{k_1 \ldots k_n} A_{\tilde{k}_1 \ldots \tilde{k}_n} 
\bigg( c^\dagger_{k_1}  d^\dagger_{\tilde{k}_1}
- \frac{n \eta}{4}(d_{k_1}^{\dagger}d_{\tilde{k}_1}^{\dagger}	 +c_{\tilde{k}_1}^{\dagger}c_{k_1}^{\dagger}) \bigg) c^\dagger_{k_2}  \dots c^\dagger_{k_n}   d^\dagger_{\tilde{k}_2} \dots d^\dagger_{\tilde{k}_n}  \ket{0}
\end{align}
To compute the corrections to the QFI, consider
\begin{align}
\ket{\dot{\psi}_{\varphi}^{(0)}} \approx &  \multInt \frac{\prod_{i,j=1}^{n} \dd k_i \dd \tilde{k}_j }{\sqrt{1-n\eta}(2\pi)^{2n} n!^2}A_{k_1 \ldots k_n} A_{\tilde{k}_1 \ldots \tilde{k}_n} 
\nonumber\\
& \quad 
\bigg( n d^\dagger_{k_1}  d^\dagger_{\tilde{k}_1} c^\dagger_{k_2}  d^\dagger_{\tilde{k}_2} 
+n c^\dagger_{k_1}  c^\dagger_{\tilde{k}_1}c^\dagger_{k_2}  d^\dagger_{\tilde{k}_2}
-\frac{n \eta}{4}\bigg[2c_{k_1}^{\dagger}d_{\tilde{k}_1}^{\dagger}	c^\dagger_{k_2}  d^\dagger_{\tilde{k}_2} 
+2n d_{k_1}^{\dagger}c_{\tilde{k}_1}^{\dagger}c^\dagger_{k_2}  d^\dagger_{\tilde{k}_2} 
\nonumber\\
& \quad 
+(n-1) d_{k_1}^{\dagger}d_{\tilde{k}_1}^{\dagger}d^\dagger_{k_2}  d^\dagger_{\tilde{k}_2}
+(n-1) c_{k_1}^{\dagger}c_{\tilde{k}_1}^{\dagger}c^\dagger_{k_2}  c^\dagger_{\tilde{k}_2}\bigg]
\bigg)  
c^\dagger_{k_3}  \dots c^\dagger_{k_n}   d^\dagger_{\tilde{k}_3} \dots d^\dagger_{\tilde{k}_n} \ket{0}.
\end{align}
Using the short-hand notation
\begin{equation}
\int \dd X f(X) 
\equiv \multInt \frac{\prod_{i,j}^{n} \dd k_i \dd \tilde{k}_j \dd p_i \dd \tilde{p}_j}{(2\pi)^{4n} n!^4} 
A_{p_1 \cdots p_n}^* A_{\tilde{p}_1 \cdots \tilde{p}_n}^* A_{k_1 \cdots k_n} A_{\tilde{k}_1 \cdots \tilde{k}_n} f(X).
\end{equation}
we proceed to compute
\begin{align}
\braket{\psi_{\varphi}^{(0)}}{\dot{\psi}_{\varphi}^{(0)}}&\approx \frac{-n\eta}{4} \bigg(
\int \dd X \bra{0}c_{p_1} \cdots c_{p_n}d_{\tilde{p}_1} \cdots d_{\tilde{p}_n} \left(2c_{k_1}^{\dagger}d_{\tilde{k}_1}^{\dagger}	c^\dagger_{k_2}  d^\dagger_{\tilde{k}_2} 
+2n d_{k_1}^{\dagger}c_{\tilde{k}_1}^{\dagger}c^\dagger_{k_2}  d^\dagger_{\tilde{k}_2} \right) c^\dagger_{k_3}  \dots c^\dagger_{k_n}   d^\dagger_{\tilde{k}_3} \dots d^\dagger_{\tilde{k}_n}\ket{0}
\nonumber\\
&\quad+n\int \dd X \bigg[\bra{0}c_{p_1} \cdots c_{p_n}c_{\tilde{p}_1} d_{\tilde{p}_2} \cdots d_{\tilde{p}_n}
c^\dagger_{k_1}  \dots c^\dagger_{k_n} c^\dagger_{\tilde{k}_1}  d^\dagger_{\tilde{k}_2} \dots d^\dagger_{\tilde{k}_n}\ket{0}
+c\leftrightarrow d\bigg]\bigg)
\nonumber\\
&=\frac{-n\eta}{4} \bigg(2+2nI_{2n}
+n\int \dd X \bigg[\bra{0}c_{p_1} \cdots c_{p_n}c_{\tilde{p}_1} d_{\tilde{p}_2} \cdots d_{\tilde{p}_n}
c^\dagger_{k_1}  \dots c^\dagger_{k_n} c^\dagger_{\tilde{k}_1}  d^\dagger_{\tilde{k}_2} \dots d^\dagger_{\tilde{k}_n}\ket{0}
+c\leftrightarrow d\bigg]\bigg),
\end{align}
and the second term yields,
\begin{align}
&\int \dd X \bra{0}c_{p_1} \cdots c_{p_n}c_{\tilde{p}_1} d_{\tilde{p}_2} \cdots d_{\tilde{p}_n}
c^\dagger_{k_1}  \dots c^\dagger_{k_n} c^\dagger_{\tilde{k}_1}  d^\dagger_{\tilde{k}_2} \dots d^\dagger_{\tilde{k}_n}\ket{0}
\nonumber\\
&=\int \dd X \bigg( [c_{\tilde{p}_1},c^{\dagger}_{\tilde{k}_1}](n-1)!\prod_{i=2}^{n} [d_{\tilde{p}_i},c^{\dagger}_{\tilde{k}_i}]  n!\prod_{i=1}^{n} [c_{p_i},c^{\dagger}_{k_i}] 
+ 
n[c_{\tilde{p_1}},c^{\dagger}_{k_1}] n[c_{p_1},c^{\dagger}_{\tilde{k}_1}](n-1)!\prod_{i=2}^{n} [d_{\tilde{p}_i},c^{\dagger}_{\tilde{k}_i}]  (n-1)!\prod_{i=2}^{n} [c_{p_i},c^{\dagger}_{k_i}] 
\bigg)
\nonumber\\
&=\frac{1}{n}+I_{2n},
\end{align}
putting everything together,
\begin{align}
\braket{\psi_{\varphi}^{(0)}}{\dot{\psi}_{\varphi}^{(0)}}&= -\frac{N}{2}\eta\left(1+\frac{N}{2}I_N\right)+\mathcal{O}(\eta^2)
\end{align}
where we used that $n=N/2$. A similar derivation yields, 
\begin{align}
\braket{\dot{\psi}_{\varphi}^{(0)}}{\dot{\psi}_{\varphi}^{(0)}}&= \frac{N^2}{2}I_{N}+N+\mathcal{O}(\eta^2)
\end{align}
and hence
\begin{align}
F_Q[\psi_{\varphi}^{(0)}]\approx F_Q[\psi_{\varphi}]-\frac{N^2\eta I_{N}}{4}
\end{align}
where $F_Q[\psi_{\varphi}]$ is the QFI without losses and we considered only dominant terms in $\eta$ and $N$. Note that this is a conservative bound, since we expect the other terms in Eq.~\ref{eq:qfloss}, where more photons have been lost, to also contribute to the QFI.

\end{widetext}

\end{document}